\documentclass{aastex631}
\graphicspath{{./}{figures/}}%
\hypersetup{linkcolor=red,citecolor=blue,filecolor=cyan}%,urlcolor=magenta
\usepackage{CJK}
\usepackage{footmisc} % For reusable footnotes

\begin{document}

%\title{Magnetic Field Measurement of a Magnetic Flux Rope Using Microwave and Radio Spectral Imaging Diagnostics}
\title{Measuring the Magnetic Field of a Coronal Mass Ejection from the Low to Middle Corona}

\author[0000-0002-1810-6706]{Xingyao Chen}
\affiliation{Center for Solar-Terrestrial Research, New Jersey Institute of Technology, 323 Martin Luther King Jr. Blvd., Newark, NJ 07102-1982, USA}
\author[0000-0002-0660-3350]{Bin Chen}
\affiliation{Center for Solar-Terrestrial Research, New Jersey Institute of Technology, 323 Martin Luther King Jr. Blvd., Newark, NJ 07102-1982, USA}
\author[0000-0003-2872-2614]{Sijie Yu}
\affiliation{Center for Solar-Terrestrial Research, New Jersey Institute of Technology, 323 Martin Luther King Jr. Blvd., Newark, NJ 07102-1982, USA}
\author[0000-0002-2325-5298]{Surajit Mondal}
\affiliation{Center for Solar-Terrestrial Research, New Jersey Institute of Technology, 323 Martin Luther King Jr. Blvd., Newark, NJ 07102-1982, USA}
\author[0000-0002-8538-3455]{Muriel Zoë Stiefel}
\affiliation{University of Applied Sciences and Arts Northwestern Switzerland, Bahnhofstrasse 6, 5210 Windisch, Switzerland}
\affiliation{ETH Zürich, Rämistrasse 101, 8092 Zürich Switzerland}
\author[0000-0001-6855-5799]{Peijin Zhang}
\affiliation{Center for Solar-Terrestrial Research, New Jersey Institute of Technology, 323 Martin Luther King Jr. Blvd., Newark, NJ 07102-1982, USA}
\affiliation{Cooperative Programs for the Advancement of Earth System Science, University Corporation for Atmospheric Research, Boulder, CO, USA}
\author[0000-0003-2520-8396]{Dale E. Gary}
\affiliation{Center for Solar-Terrestrial Research, New Jersey Institute of Technology, 323 Martin Luther King Jr. Blvd., Newark, NJ 07102-1982, USA}
\author[0000-0002-2002-9180]{Säm Krucker}
\affiliation{University of Applied Sciences and Arts Northwestern Switzerland, Bahnhofstrasse 6, 5210 Windisch, Switzerland}
\affiliation{Space Sciences Laboratory, University of California, 94720 Berkeley, USA}

\author[0000-0003-2238-2698]{Marin M. Anderson}
\affiliation{Owens Valley Radio Observatory, California Institute of Technology, Big Pine, CA 93513, USA}
\affiliation{Jet Propulsion Laboratory, California Institute of Technology, Pasadena, CA 91011, USA}
\author{Judd D. Bowman}
\affiliation{School of Earth and Space Exploration, Arizona State University, Tempe, AZ 85287, USA}
\author[0000-0003-4980-2736]{Ruby Byrne}
\affiliation{Cahill Center for Astronomy and Astrophysics, California Institute of Technology, Pasadena, CA 91125, USA}
\affiliation{Owens Valley Radio Observatory, California Institute of Technology, Big Pine, CA 93513, USA}
\author{Morgan Catha}
\affiliation{Owens Valley Radio Observatory, California Institute of Technology, Big Pine, CA 93513, USA}
\author[0000-0001-7754-0804]{Sherry Chhabra}
\affiliation{Center for Solar-Terrestrial Research, New Jersey Institute of Technology, 323 Martin Luther King Jr. Blvd., Newark, NJ 07102-1982, USA}
\affiliation{George Mason University, Fairfax, VA 22030, USA}
\author{Larry D'Addario}
\affiliation{Cahill Center for Astronomy and Astrophysics, California Institute of Technology, Pasadena, CA 91125, USA}
\affiliation{Owens Valley Radio Observatory, California Institute of Technology, Big Pine, CA 93513, USA}
\author[0000-0001-5397-5969]{Ivey Davis}
\affiliation{Cahill Center for Astronomy and Astrophysics, California Institute of Technology, Pasadena, CA 91125, USA}
\affiliation{Owens Valley Radio Observatory, California Institute of Technology, Big Pine, CA 93513, USA}
\author{Jayce Dowell}
\affiliation{University of New Mexico, Albuquerque, NM 87131, USA}
% \author{Katherine Elder}
% \affiliation{School of Earth and Space Exploration, Arizona State University, Tempe, AZ 85287, USA}
\author{Gregg Hallinan}
\affiliation{Cahill Center for Astronomy and Astrophysics, California Institute of Technology, Pasadena, CA 91125, USA}
\affiliation{Owens Valley Radio Observatory, California Institute of Technology, Big Pine, CA 93513, USA}
\author{Charlie Harnach}
\affiliation{Owens Valley Radio Observatory, California Institute of Technology, Big Pine, CA 93513, USA}
\author{Greg Hellbourg}
\affiliation{Cahill Center for Astronomy and Astrophysics, California Institute of Technology, Pasadena, CA 91125, USA}
\affiliation{Owens Valley Radio Observatory, California Institute of Technology, Big Pine, CA 93513, USA}
\author{Jack Hickish}
\affiliation{Real-Time Radio Systems Ltd, Bournemouth, Dorset BH6 3LU, UK}
\author{Rick Hobbs}
\affiliation{Owens Valley Radio Observatory, California Institute of Technology, Big Pine, CA 93513, USA}
\author{David Hodge}
\affiliation{Cahill Center for Astronomy and Astrophysics, California Institute of Technology, Pasadena, CA 91125, USA}
\author{Mark Hodges}
\affiliation{Owens Valley Radio Observatory, California Institute of Technology, Big Pine, CA 93513, USA}
\author{Yuping Huang}
\affiliation{Cahill Center for Astronomy and Astrophysics, California Institute of Technology, Pasadena, CA 91125, USA}
\affiliation{Owens Valley Radio Observatory, California Institute of Technology, Big Pine, CA 93513, USA}
\author{Andrea Isella}
\affiliation{Department of Physics and Astronomy, Rice University, Houston, TX 77005, USA}
\author{Daniel C. Jacobs}
\affiliation{School of Earth and Space Exploration, Arizona State University, Tempe, AZ 85287, USA}
\author{Ghislain Kemby}
\affiliation{Owens Valley Radio Observatory, California Institute of Technology, Big Pine, CA 93513, USA}
\author{John T. Klinefelter}
\affiliation{Owens Valley Radio Observatory, California Institute of Technology, Big Pine, CA 93513, USA}
\author{Matthew Kolopanis}
\affiliation{School of Earth and Space Exploration, Arizona State University, Tempe, AZ 85287, USA}
\author[0000-0003-1226-118X]{Nikita Kosogorov}
\affiliation{Cahill Center for Astronomy and Astrophysics, California Institute of Technology, Pasadena, CA 91125, USA}
\affiliation{Owens Valley Radio Observatory, California Institute of Technology, Big Pine, CA 93513, USA}
\author{James Lamb}
\affiliation{Owens Valley Radio Observatory, California Institute of Technology, Big Pine, CA 93513, USA}
\author{Casey J Law}
\affiliation{Cahill Center for Astronomy and Astrophysics, California Institute of Technology, Pasadena, CA 91125, USA}
\affiliation{Owens Valley Radio Observatory, California Institute of Technology, Big Pine, CA 93513, USA}
\author{Nivedita Mahesh}
\affiliation{Cahill Center for Astronomy and Astrophysics, California Institute of Technology, Pasadena, CA 91125, USA}
\affiliation{Owens Valley Radio Observatory, California Institute of Technology, Big Pine, CA 93513, USA}
\author{Brian O'Donnell}
\affiliation{Center for Solar-Terrestrial Research, New Jersey Institute of Technology, 323 Martin Luther King Jr. Blvd., Newark, NJ 07102-1982, USA}
\author{Kathryn Plant}
\affiliation{Owens Valley Radio Observatory, California Institute of Technology, Big Pine, CA 93513, USA}
\affiliation{Jet Propulsion Laboratory, California Institute of Technology, Pasadena, CA 91011, USA}
\author{Corey Posner}
\affiliation{Owens Valley Radio Observatory, California Institute of Technology, Big Pine, CA 93513, USA}
\author{Travis Powell}
\affiliation{Owens Valley Radio Observatory, California Institute of Technology, Big Pine, CA 93513, USA}
\author{Vinand Prayag}
\affiliation{Owens Valley Radio Observatory, California Institute of Technology, Big Pine, CA 93513, USA}
\author{Andres Rizo}
\affiliation{Owens Valley Radio Observatory, California Institute of Technology, Big Pine, CA 93513, USA}
\author{Andrew Romero-Wolf}
\affiliation{Jet Propulsion Laboratory, California Institute of Technology, Pasadena, CA 91011, USA}
\author{Jun Shi}
\affiliation{Cahill Center for Astronomy and Astrophysics, California Institute of Technology, Pasadena, CA 91125, USA}
\author{Greg Taylor}
\affiliation{University of New Mexico, Albuquerque, NM 87131, USA}
\author{Jordan Trim}
\affiliation{Owens Valley Radio Observatory, California Institute of Technology, Big Pine, CA 93513, USA}
\author{Mike Virgin}
\affiliation{Owens Valley Radio Observatory, California Institute of Technology, Big Pine, CA 93513, USA}
\author{Akshatha Vydula}
\affiliation{School of Earth and Space Exploration, Arizona State University, Tempe, AZ 85287, USA}
\author{Sandy Weinreb}
\affiliation{Cahill Center for Astronomy and Astrophysics, California Institute of Technology, Pasadena, CA 91125, USA}
% \author{Scott White}
% \affiliation{Owens Valley Radio Observatory, California Institute of Technology, Big Pine, CA 93513, USA}
\author{David Woody}
\affiliation{Owens Valley Radio Observatory, California Institute of Technology, Big Pine, CA 93513, USA}
% \author{Thomas Zentmeyer}
% \affiliation{Owens Valley Radio Observatory, California Institute of Technology, Big Pine, CA 93513, USA}

\correspondingauthor{Xingyao Chen}
\email{xingyao.chen@njit.edu}
\correspondingauthor{Bin Chen}
\email{bin.chen@njit.edu}

%%%%%%%%%%%%%%%%%%%%%%%%%%%%%%%%%%%%%%%%%%%%%%%%%%
\begin{abstract}
A major challenge in understanding the initiation and evolution of coronal mass ejections (CMEs) is measuring the magnetic field of the magnetic flux ropes (MFRs) that drive CMEs.
Recent developments in radio imaging spectroscopy have paved the way for diagnosing the CMEs' magnetic field using gyrosynchrotron radiation. 
We present magnetic field measurements of a CME associated with an X5-class flare by combining radio imaging spectroscopy data in microwaves (1--18 GHz) and meter-wave (20--88 MHz), obtained by the Owens Valley Radio Observatory's Expanded Owens Valley Solar Array (EOVSA) and Long Wavelength Array (OVRO-LWA), respectively. 
EOVSA observations reveal that the microwave source, observed in the low corona during the initiation phase of the eruption, outlines the bottom of the rising MFR-hosting CME bubble seen in extreme ultraviolet and expands as the bubble evolves. 
As the MFR erupts into the middle corona and appears as a white light CME, its meter-wave counterpart, observed by OVRO-LWA, displays a similar morphology. 
For the first time, using gyrosynchrotron spectral diagnostics, we obtain magnetic field measurements of the erupting MFR in both the low and middle corona, corresponding to coronal heights of 0.02 and 1.83 $R_{\odot}$. 
The magnetic field strength is found to be around 300 G at 0.02 $R_{\odot}$ during the CME initiation, and about 0.6 G near the leading edge of the CME when it propagates to 1.83 $R_{\odot}$.
These results provide critical new insights into the magnetic structure of the CME and its evolution during the early stages of its eruption.
\end{abstract}

%%%%%%%%%%%%%%%%%%%%%%%%%%%%%%%%%%%%%%%%%%%%%%%%%%MFR magnetic field firstly
\section{Introduction}
\label{sec-intro}

Coronal mass ejections (CMEs) are believed to be driven by magnetic flux ropes (MFRs), which are three-dimensional helical magnetic structures characterized by magnetic field lines twisted around a central axis \citep{2000ApJ...545..524C, 2001ApJ...559..452Z, 2011LRSP....8....1C, 2017ScChD..60.1383C, 2020RAA....20..165L}. 
%%==========to measure B evolution of MFR
Owing to the key role that MFRs play in the development and evolution of CMEs, it is imperative to obtain detailed knowledge of their magnetic field. However, due to the lack of observational constraints on the magnetic field in the solar corona, direct measurements of MFRs' magnetic field remain rare. Previously, the knowledge of MFRs' magnetic field configuration during their pre-eruption phase near the solar surface has primarily relied on magnetic field extrapolation techniques based on spectropolarimetry measurements taken from the photosphere \citep{2014ApJ...786L..16J, 2015NatCo...6.7008W, 2021NatCo..12.2734Z, 2023ApJ...951...54Y}. 
Alternatively, magnetohydrodynamic simulations have been applied to model the evolution of MFRs by incorporating photospheric magnetogram measurements \citep[see, e.g., a recent review by][and references therein]{2022Innov...300236J}.
As the MFR propagates into interplanetary space, occasionally, its magnetic field can be directly measured by \textit{in situ} instruments, manifesting itself with a characteristic rotation of the magnetic field vector as the MFR passes by from the spacecraft \citep{2014ApJ...793...53H, 2019ApJ...877...77V}. 

%%==========gyrosynchrotron from MW and radio 
%MW emission in radio to deduce the B
Gyrosynchrotron radiation, which occurs when high-energy electrons spiral around magnetic field lines, provides an excellent means of constraining the coronal magnetic field \citep{1985ARA&A..23..169D, 1998ARA&A..36..131B, 2023ARA&A..61..427G}. The intensity and spectral characteristics of these emissions depend on the energy distribution of the electrons, the magnetic field strength, and the viewing angle. Owing to the different magnetic field environments, gyrosynchrotron radiation usually peaks in the microwave regime for flares and CMEs occurring in the low corona and in meter-decameter waves for events in the middle corona (loosely defined as a radial distance of $\sim\,1.5-6R_{\odot}$; \citealt{West2023}). For example, using microwave imaging spectroscopy observations obtained by the Expanded Owens Valley Solar Array (EOVSA; \citealt{Gary2018}), \citet{2020ApJ...895L..50C} revealed microwave emission from conjugate footpoints of a magnetic flux rope during the early energy release phase of the associated X8.2-class eruptive flare. The magnetic field strength there was derived to be 300--500 G. 
As the MFR propagates into the middle corona with smaller magnetic field strength, gyrosynchrotron radiation peaks at longer, meter-decameter wavelengths. Spatially resolved metric spectral observations of CME-associated gyrosynchrotron radiation have served as a powerful tool for constraining the magnetic fields at different parts of the CME \citep{2001ApJ...558L..65B, 2007ApJ...660..874M,2013ApJ...766..130T, 2014ApJ...782...43B, 2017A&A...608A.137C, 2020ApJ...893...28M, Chhabra2021, 2023ApJ...950..164K}. 
Recently, using data obtained by the Murchison Widefield Array (MWA), \citet{2024ApJ...968...55K} detected gyrosynchrotron radiation from a CME out to 8.3 $R_{\odot}$, with magnetic field derived to $\sim$3 $R_{\odot}$ with imaging spectropolarimetry.

In this Letter, we present, for the first time, measurements of the magnetic field of an erupting magnetic flux rope during its initiation and propagation phases at low- and middle-coronal heights of 0.02 and 1.83 $R_{\odot}$, respectively. These measurements are achieved by utilizing nonthermal gyrosynchrotron spectral diagnostics based on microwave and metric wavelength spectral imaging observations obtained by EOVSA and the newly commissioned Owens Valley Radio Observatory's Long Wavelength Array\footnote{An instrumental paper describing the OVRO-LWA is currently under preparation.} (OVRO-LWA; see \citealt{2018ApJ...864...22A} for general descriptions of the instrument), respectively, with support from complementary multi-wavelength data from a variety of instruments. 
Section \ref{sec-obs} details the observations and data analysis. Section \ref{sec-dis} discusses and summarizes the main findings.

%%%%%%%%%%%%%%%%%%%%%%%%%%%%%%%%%%%%%%%%%%%%%%%%%%

%%==========Figures
\begin{figure*}%[!ht]
\centering
\includegraphics[width=18cm]{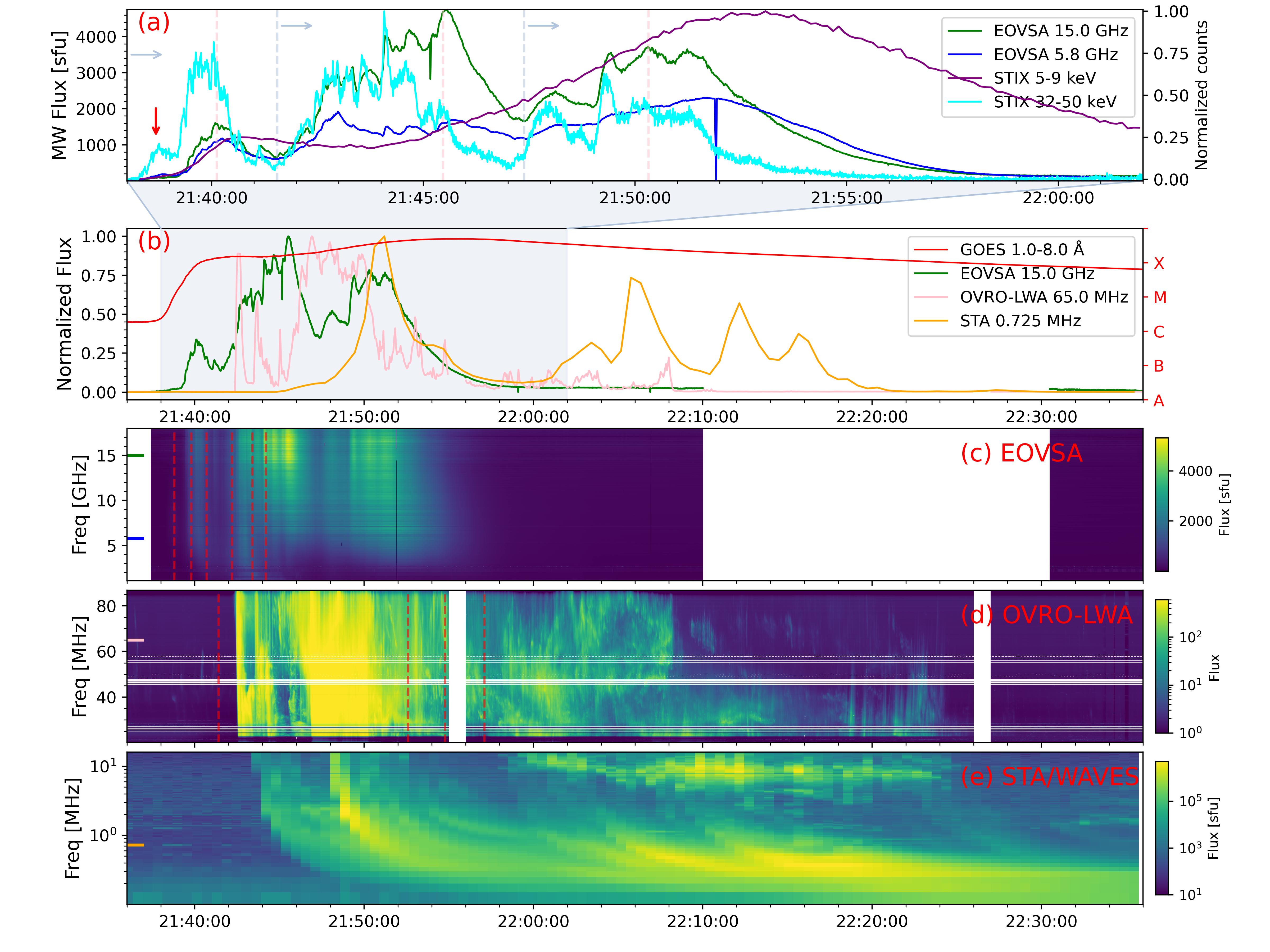}
\caption{Time profiles of: (a) EOVSA microwave emissions at 5.8 and 15.0 GHz, and STIX X-ray emissions in the energy bands of 5-9 keV and 32-50 keV; (b) GOES soft X-ray flux, EOVSA microwave emissions at 15 GHz, OVRO-LWA metric wave emissions at 65 MHz, and STEREO-A kilometric emissions at 0.725 MHz. 
Dynamic spectra observed by: (c) EOVSA (1-18 GHz), (d) OVRO-LWA (20-87 MHz), and (e) STEREO-A (0.125-16.025 MHz). 
In the dynamic spectra, the red dashed vertical lines indicate the imaging times for microwave (Figure~\ref{fig3}) and metric radio emissions (Figure~\ref{fig5}), while the horizontal short lines mark the frequencies corresponding to the flux curves, with the specified colors.
The dashed lines in panel (a) mark three peak and two valley times for microwave light curves at 15 GHz.
The STIX flux in 5-9 keV is taken from the background detector.%to have no attenuation.
\label{fig1}}
\end{figure*}

%%==========Figures
\begin{figure*}%[!ht]
\centering
\includegraphics[width=18cm]{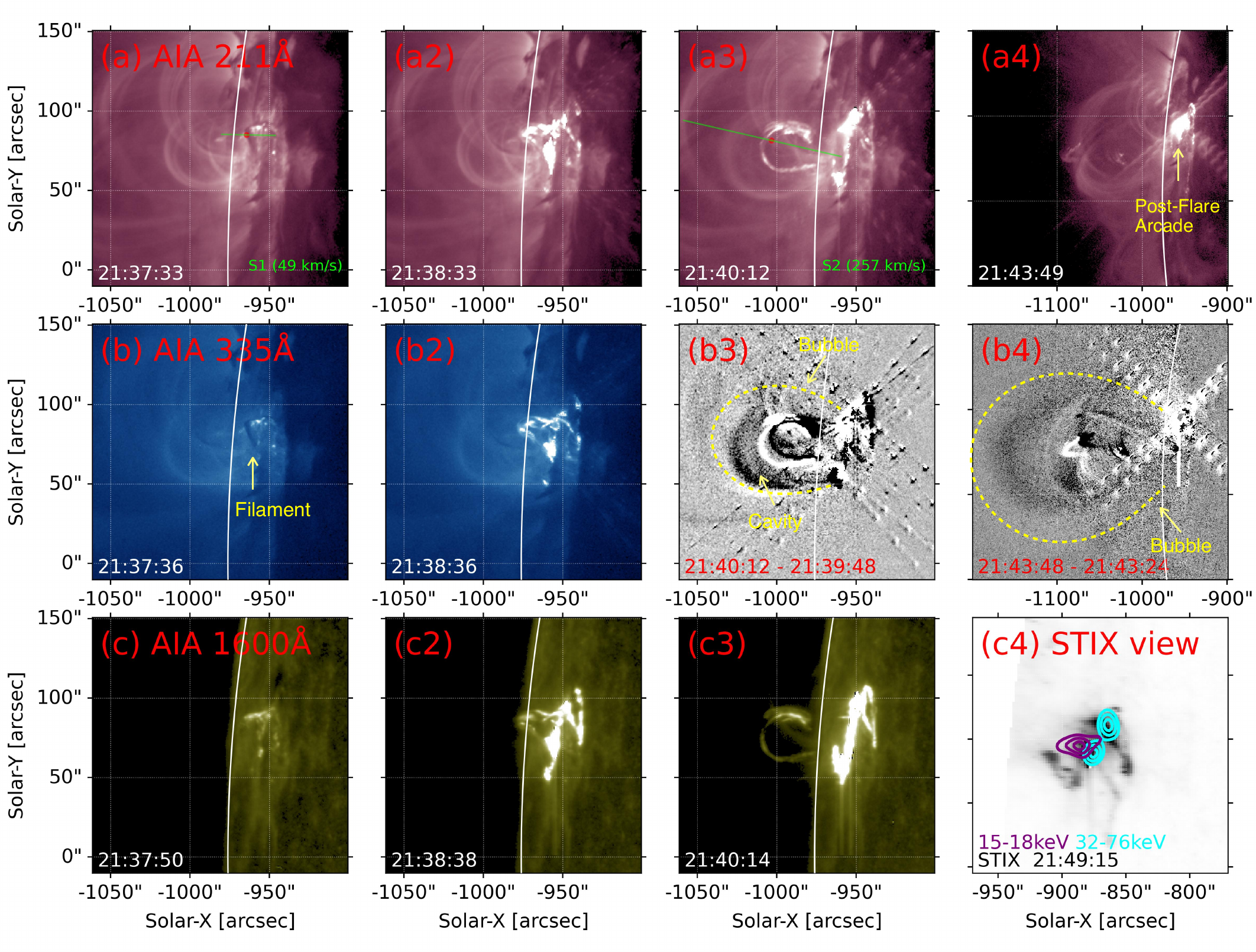}
\caption{EUV, and X-ray images at various times related to the X5 flare, including: 
(a1-a4) AIA 211 $\AA$, (b1-b4) AIA 335 $\AA$, (c1-c3) AIA 1600 $\AA$, and (c4) STIX X-ray images.
The X-ray images are shownin the energy bands of 15-18 keV and 32-76 keV, integrated over the time interval 21:49:15-21:50:15, with contours at 50\%, 70\%, and 90\% of the peak count rate, overlaid on the AIA 1600 $\AA$ map as projected to the SolO/STIX perspective, with an $x$-range of (-970$''$, -770$''$) and $y$-range of (-55$''$, 145$''$).
%The integration times of STIX images are 21:38:25-21:39:10, 21:39:35-21:40:00, and 21:43:35-21:44:00, respectively for the last three columns.
% The times of the STIX images are indicated, in the last row, with each image integrated over 20 seconds.
%21:39:15-21:39:35, 21:40:25-21:40:45, 21:42:25-21:42:45, and 21:49:15-21:49:35, 
The last two panels in (b) AIA 335~$\AA$ are shown as running-difference images, computed by subtracting the image at 24 seconds earlier from the current time.
In panel (a), green lines represent two slices (S1 and S2) used for the time-distance plot, from which the gradual rise speed of the filament was derived as 49 km/s between 21:36 and 21:38 UT, and the ascending velocity of the erupting bubble was measured at 257 km/s between 21:38 and 21:43 UT.
\label{fig2}}
\end{figure*}

%%==========Figures
\begin{figure}%[!ht]
\centering
\includegraphics[width=8cm]{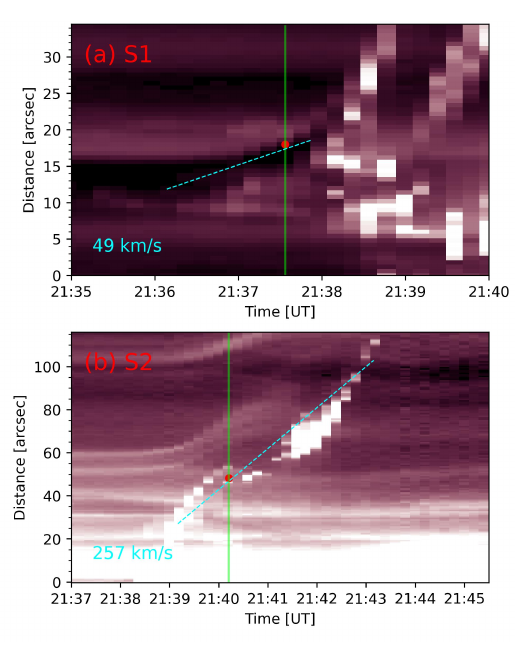}
\caption{Time--distance plots along two slices in the AIA 211 Å images, corresponding to Slice 1 and Slice 2 as indicated in Figure~\ref{fig2}(a).
\label{fig2s}}
\end{figure}

%%==========Figures
\begin{figure*}%[!ht]
\centering
\includegraphics[width=18cm]{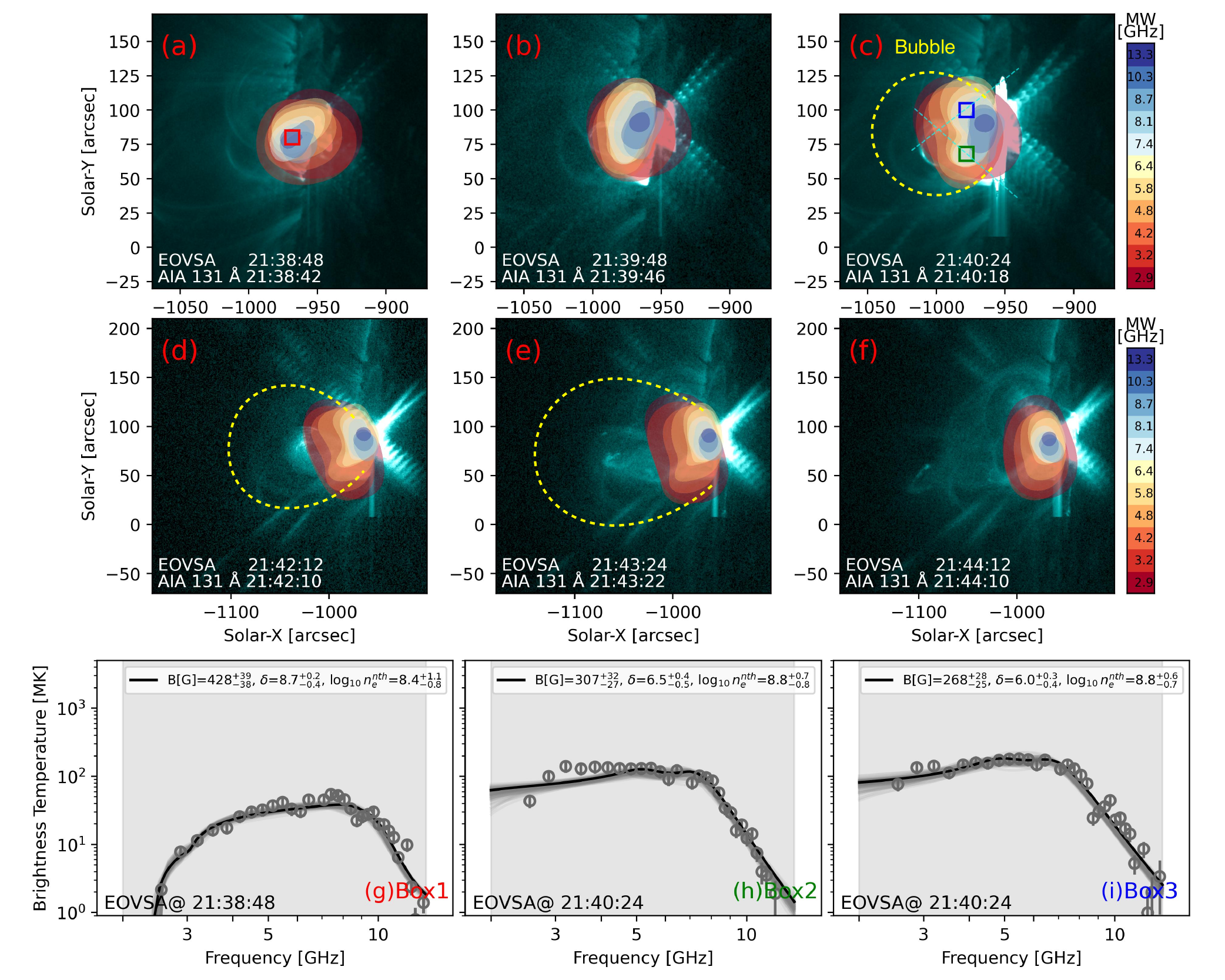}
\caption{EOVSA imaging and spectral fitting results.
(a)-(f) EOVSA images at multiple frequencies and times with contours of 50\% peak brightness temperature; (g), (h), and (i) present the microwave spectral fit results for the regions corresponding to the three distinct boxes marked in panels (a) and (c).
The dashed lines in (c), (d), and (e) outline the edges of the bubble.
The dashed thin lines in cyan color in (c) indicate the slices to estimate the cross-section area of the MFR legs.
The fitted parameters, including the magnetic field strength ($B$), power-law index ($\delta$), and non-thermal electron number density ($n_e^{nth}$), are indicated in the legends for each spectral fit.
\label{fig3}}
\end{figure*}

%%==========Figures
\begin{figure*}%[!ht]
\centering
\includegraphics[trim={1cm 0 0 0},clip,width=18cm]{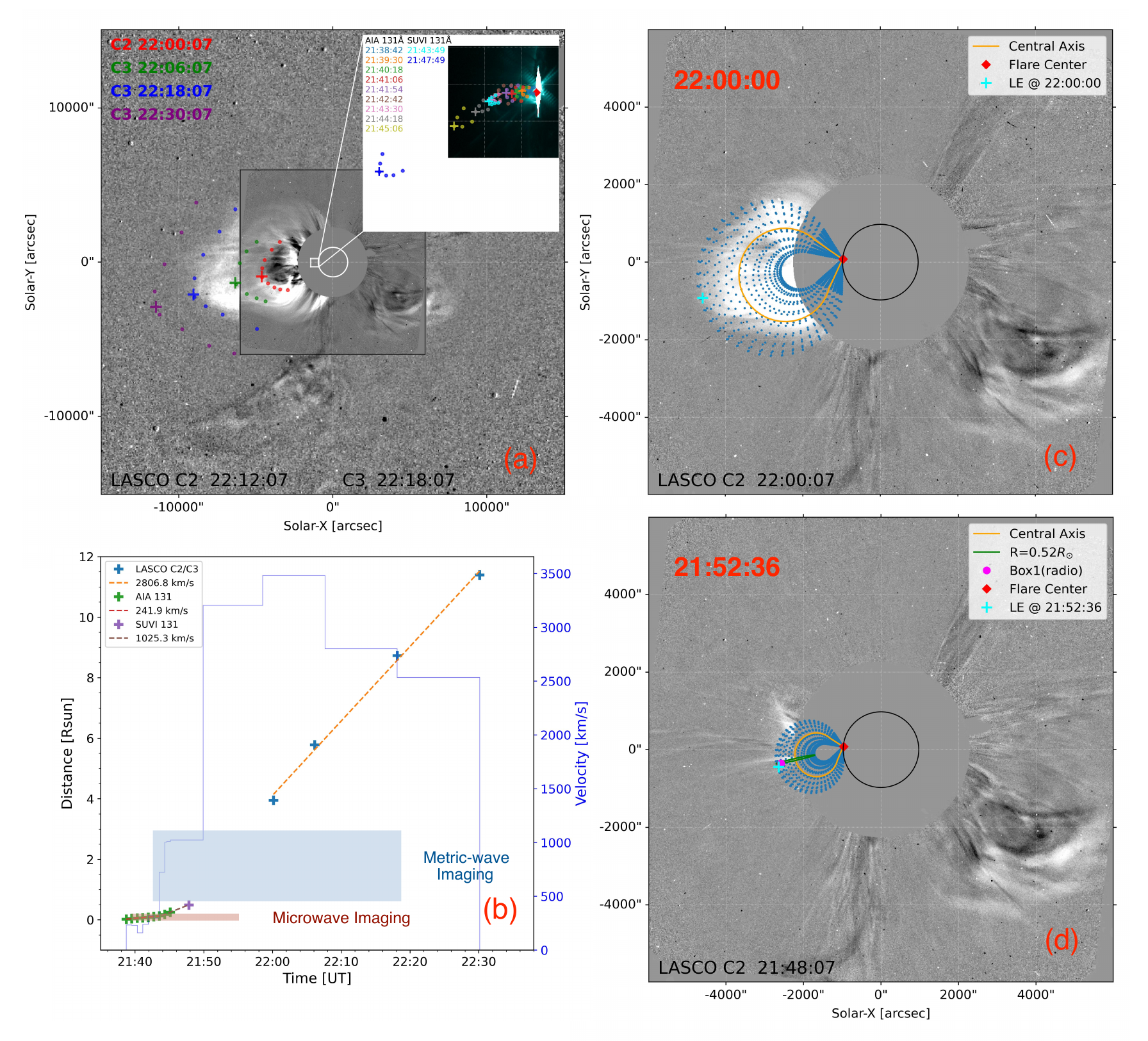}
\caption{(a) Trajectory of the leading edge of filament at different times, observed by AIA and SUVI in the 131 \AA band, and leading edge of the erupting bubble as seen in LASCO C2 (22:12:05–22:00:05 UT) and C3 (22:18:05–22:06:05 UT). The dots outline the leading edge, while the plus signs mark the farthest position of the leading edge from the flare center.
(b) Time-varying distance between the leading edge and the flare center, corresponding to the path in plus signs in panel (a). The indicated speed is derived from a linear fitting of the data. The shaded red and blue regions represent the observed times and the heliocentric distances from microwave and metric radio imaging sources, respectively.
Another approach is applied to estimate the velocity of the moving front by using the difference in its position over time, as indicated by the blue stairs on the secondary y-axis.
(c)-(d) Reconstruction of the CME using GCS model at 22:00:00 UT (c), showing the MFR bubble as firstly observed in the LASCO C2 field of view, and 21:52:36 UT (d), marking the time of metric-wave imaging and spectral analysis.
%(e) Meter-wavelength radio sources at 34.1 MHz (purple) and 80.0 MHz (green) from OVRO-LWA, alongside microwave sources at 4.8 GHz (orange) and 7.4 GHz (blue) from EOVSA, overlaid on AIA 131 \AA image and the time-difference image from LASCO C2. The trajectory is indicated by the red dashed line, with the flare center marked by a red diamond symbol.
%(f) Radio sources at 34.1 MHz and 80.0 MHz are shown to follow the CME path, shaping the CME morphology at 21:55 UT.
\label{fig4}}
\end{figure*}

%%==========Figures
\begin{figure*}%[!ht]
\centering
\includegraphics[width=18cm]{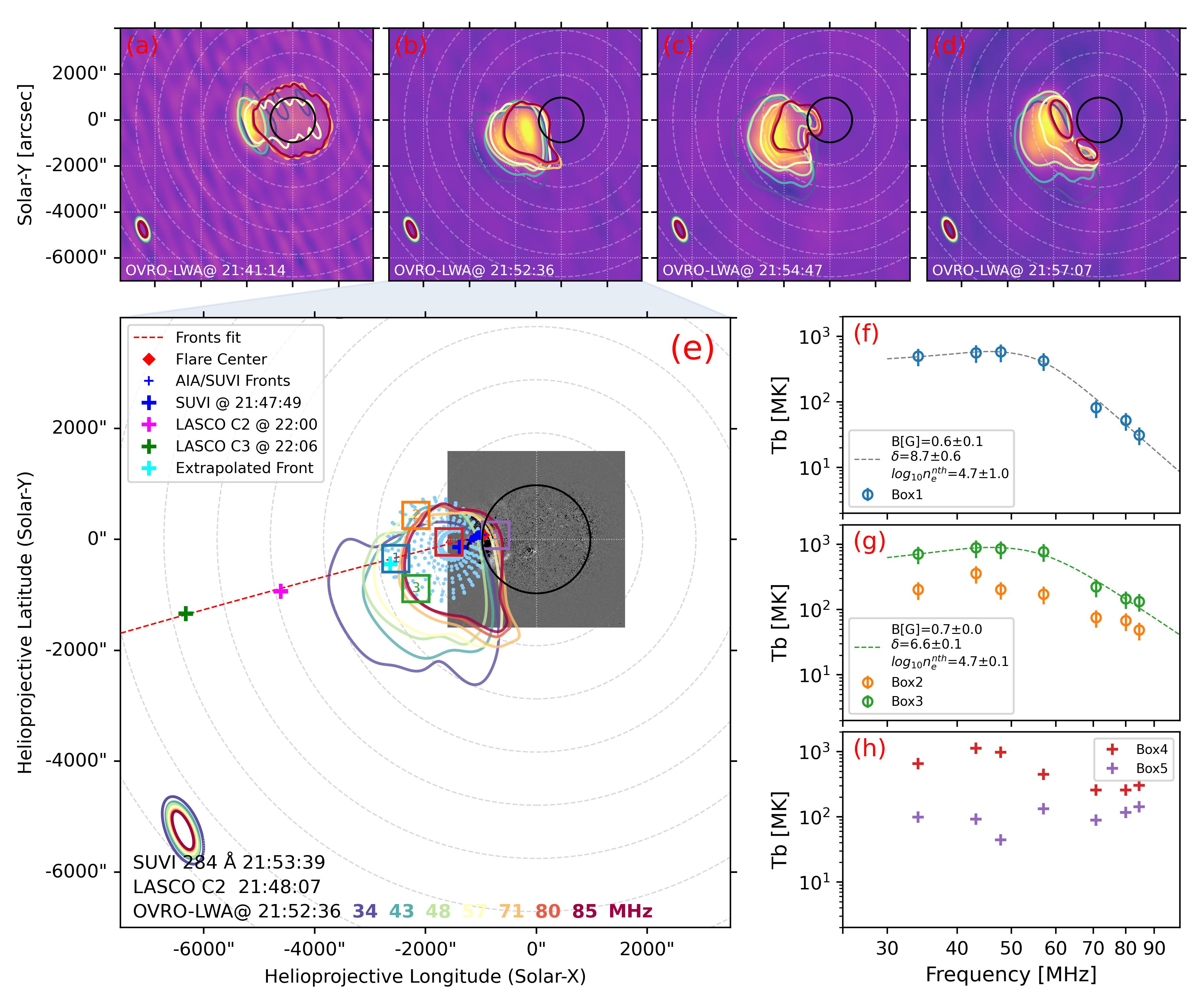}
\caption{Metric-wave radio imaging and spectral fitting results.
(a)-(d) OVRO-LWA images at 21:41:14 UT, 21:52:36 UT, 21:54:47 UT, and 21:57:07 UT, respectively. The background in each panel shows the 48 MHz image, with overlaid contours at 34, 43, 48, 57, 71, 80, and 85 MHz, plotted at 10\% of the peak brightness temperature.  The corresponding elliptical beam size and orientation are shown in the lower left corner of each panel.
(e) Radio image used for spectral fitting at 21:52:36 UT, overlaid with contours at 20\% of the peak brightness temperature.
The grayscale background image is the SUVI time-difference image, which is derived by subtracting the previous closest time-cadence image.
Leading edges identified from SUVI and LASCO are marked with plus signs. The trajectory of the MFR leading edge is shown as a red dashed line. The cyan plus sign indicates the extrapolated leading edge position at 21:52:36 UT, as there were no direct imaging observations of the MFR available from SUVI or LASCO at this time. Three boxes are defined to obtain the averaged brightness temperature.
(f)-(h) OVRO-LWA metric-wave spectral light curves for the five boxes. GS spectral fitting results are shown for box 1 and box 3.
Panels (a)–(e) have the same field of view in Solar-X and Solar-Y coordinates.
\label{fig5}}
\end{figure*}

%%%%%%%%%%%%%%%%%%%%%%%%%%%%%%%%%%%%%%%%%%%%%%%%%%
\section{Observations}
\label{sec-obs}
%\subsection{Overviews}
%\label{sec-obs-overview}

The CME event under study is associated with an X5-class eruptive solar flare that occurred on December 31, 2023. The flare starts at 21:36 UT, peaks at 21:55 UT, and ends at 22:13 UT. The source region is NOAA active region (AR) 13536 located around N04E79. It is followed by a white light CME which first appeared in the LASCO C2's field of view at 22:00, according to the LASCO CME CATALOG
%\footnote{\label{cme_catalog}\url{https://cdaw.gsfc.nasa.gov/CME_list/}}
\footnote{\label{cme_catalog}https://cdaw.gsfc.nasa.gov/CME\_list}.
%, which is the most powerful flare since X8.2-class flare on September 10, 2017.
%A CME firstly appeared in the LASCO C2 view at 22:00 with a speed of 2852 km/s from the linear fit to the height-time measurements from SOHO LASCO CME CATALOG.
%https://cdaw.gsfc.nasa.gov/CME_list/catalog_description.htm
%of 548 km/s and in a range of 219 to 2016 km/s from CACTus catalogue. https://www.sidc.be/cactus/catalog/LASCO/2_5_0/cme_qkl.txt

The event was observed in microwaves by EOVSA in 1--18 GHz and in meter-decameter waves by OVRO-LWA in 20--88 MHz, as well as in X-rays by the Spectrometer Telescope for Imaging X-rays on Solar Orbiter \citep[STIX/SOLO;][]{2020A&A...642A..15K}, along with other instruments including extreme ultra-violet (EUV) data from the Atmospheric Imaging Assembly/Solar Dynamics Observatory \citep[AIA/SDO;][]{2012SoPh..275...17L} and the Solar Ultraviolet Imager \citep[SUVI;][]{2018ApJ...852L...9S} on board GOES-R, the kilometer-wavelength radio flux from WAVES on Solar Terrestrial Relations Observatory \citep[WAVES/STEREO;][]{2008SSRv..136..487B} and white light by the Large Angle and Spectrometric Coronagraph on board the Solar and Heliospheric Observatory \citep[LASCO/SOHO;][]{1995SoPh..162..357B}.

%%%%%%%%%%%%%%%%%%%%%%%%%%%%%%%%%%%%%%%%%%%%%%%%%%
\subsection{Time history and radio dynamic spectra}
\label{sec-obs-flux}

The time profiles of GOES X-ray, EOVSA microwave, STIX X-ray, OVRO-LWA and STA/WAVES radio emissions are presented in Figures~\ref{fig1}(a) and (b), while the dynamic spectra for microwave, metric-wavelength, and kilometric wavelength emissions are displayed in Figures~\ref{fig1}(c), (d), and (e).
%%%%%MW
The microwave emission shown in Figure~\ref{fig1}(c) exhibits a flux increase from 21:38:25 UT during the early impulsive phase. 
Three distinct episodes (approximately marked by the light blue arrows) can be identified around 15 GHz, with corresponding peaks (dashed pink vertical lines) and valleys (dashed light blue vertical lines), as shown in Figure~\ref{fig1}(a).
%Three distinct episodes can be identified at around 15 GHz, with peak times occurring at 21:40:07 UT, 21:45:28 UT, and 21:50:19 UT,}
%and two valleys, at times of 21:41:33 UT, 21:47:23 UT as indicated in the dashed vertical lines in Figure~\ref{fig1}(a). 
After the first impulsive phase of the light curve at 15 GHz, the second phase produces the most intense microwave emission, followed by decay and subsequent rise leading to the third peak.
%Together with the presence of multiple peaks superimposed within the three phases, it indicates intermittent episodes of energy release and particle acceleration occurring throughout the flare's evolution.
%The microwave flux at 15 GHz consistently being larger than the flux at 5.8 GHz may suggest that the accelerated electrons are highly energetic, emitting more power at higher frequencies. This also indicates the presence of multiple episodes of magnetic reconnection and energy release processes.

%%%%%STIX
Likewise, the STIX 32--50 keV hard X-ray (HXR) emission, shown in Figure~\ref{fig1}(a) as the cyan curve, exhibits multiple peaks and can be grouped into three main episodes that correlate temporally with the microwave emission, suggesting a synchronized emission process from flare-accelerated energetic electrons.
%\textbf{[needed to be confirmed and re-discribe how to get the unattenuated light curve, use the detector with a pinhole?] The count rates profiles reveal that below 25 keV, saturation was observed from 21:40:10 to 22:00:30. Count rates at lower energy bands are attenuated and autonomously invoked by on-board Rate Control Regime algorithm in order to prevent detector saturation, which can be re-calculated from the detector response matrix \citep{2020A&A...642A..15K}. The count rates in 5-9 keV shown in Figure \ref{fig1} (a) are corrected from the detector attenuation.}
The 5--9 keV soft X-ray (SXR) light curve, also measured by STIX, displays the thermal emission of the flare and follows a similar shape as seen by the GOES 1--8 \AA\ SXR channel in Figure~\ref{fig1}(b). Because the attenuator moved in for the flare, which led to the attenuation of low X-ray energies below \textasciitilde11 keV \citep{2020A&A...642A..15K}, this light curve is taken from the background detector of STIX, which is the only detector not covered by the attenuator \citep{stiefel2025}.
%The flare triggered the attenuator to move in at around 21:40:10, which led to the attenuation of low X-ray energies below \textasciitilde11 keV \citep{2020A&A...642A..15K}. Therefore, the light curve from 5-9 keV is taken from the background (BKG) detector of STIX which is the only detector not covered by the attenuator \citep{stiefel2025}. 
%The mask of the BKG detector are six openings with different sizes, always two have the same size for redundandency.  The lightcurves shown in Figure~\ref{fig1} (a) are measured from the BKG detector pixels behind the middle sized openings.

%%%%%LWA
Figures~\ref{fig1}(d) and (e) show the meter-, decameter, and kilo-wave dynamic spectrum obtained by OVRO-LWA and STEREO-A/WAVES, respectively. The OVRO-LWA dynamic spectrum is recorded with its beamforming mode, with time and frequency resolutions of 64 ms and 24 kHz, respectively. The STEREO-A/WAVES dynamic spectrum has a time resolution of 35 seconds across a frequency range from 0.125 to 16.025 MHz, with a frequency resolution of 50 kHz. The intensity has been corrected to 1 AU values.
Despite the meter-decameter dynamic spectrum displaying a variety of fine structures and spiky temporal profiles, the overall emission is delineated into three distinct phases. Initially, there is a rapid increase in intensity at 21:42:22 in the OVRO-LWA dynamic spectrum. The rapid drift feature of the initial radio burst suggests that it can be characterized as a type III radio burst. %This is followed by striae and a reverse type III burst pair with short frequency bandwidths in the initial phase. 
%\textbf{During the impulsive phase, there is pronounced broadband emission along with complex substructures. Subsequently, the intensity gradually decreases, and a type II burst along with a type IV-like burst is observed following the impulsive phase.}
During the second phase around 21:50 UT, there is a pronounced broadband emission along with complex substructures in the OVRO-LWA dynamic spectrum. In the lower-frequency STEREO-A/WAVES data, this phase is featured by multiple type III bursts. Starting at around 21:55 UT, the OVRO-LWA spectrum shows a complex type II-like burst, with its low-frequency leading edge continuing into the STEREO-A/WAVES frequency range. The presence of the type II burst is indicative of the presence of a coronal shock during this period, likely driven by the accompanying CME. At lower frequencies, multiple type III radio bursts are observed, suggesting continuing injection of electron beams into the upper corona. %Subsequently, the intensity gradually decreases, and a type II burst along with a type IV-like burst is observed following the second phase.

\subsection{The Erupting Flux Rope in the Low Corona}
\label{sec-obs-euv}

\subsubsection{EUV and X-ray Observations}
%%%%%flare onset, filament
The onset of the flare is characterized by the initial ascent of a filament, visible in the SDO/AIA 211 \AA\ (sensitive to temperatures around 2 MK) channel as a dark feature (Figure~\ref{fig2}(a)). To evaluate the kinematics of the erupting filament, two slices were made in the 211 \AA\ image (green lines in Figures~\ref{fig2}(a) labeled ``S1'' and ``S2'') to produce time-distance plots. The velocities are estimated by performing linear fitting of the moving dark feature from the time-distance plots. This filament exhibits a gradual rise at a speed of 49 km/s from 21:36 to 21:38 UT, seen from the height-time map in Figure \ref{fig2s}(a), which coincides in time with the HXR precursor at 21:38:40 UT prior to the first main HXR peak (red arrow in Figure~\ref{fig1}(a)). Concurrently, a pair of flare ribbons is observed, a phenomenon typically attributed to magnetic reconnection processes triggered by the eruption of a filament-hosting MFR.

%%%%%erupting bubble
After 21:38 UT, the filament undergoes noticeable acceleration, heating, and expansion, continuing its ascent in conjunction with a plasma bubble best seen in SDO/AIA 335 \AA\ EUV channel as the MFR structure pushing outward the surrounding coronal plasma (outlined in Figure~\ref{fig2}(b)). The yellow dashed outline roughly indicates the outer boundary of the erupting MFR.
The ascending velocity of the filament is estimated at approximately 257 km/s between 21:38 and 21:43 UT, seen from the height-time map in Figure \ref{fig2s}(b).
The filament and the plasma bubble are believed to be different manifestations of the same erupting MFR system, with the former representing the dense material supported by concave-upward magnetic field lines near the bottom of the MFR and the latter outlining the volumetric MFR body \citep[see, e.g., a review by][]{2011LRSP....8....1C}.
The central cavity is seen as a low-density, dark region appearing at the core of the flux rope, as pointed out in the third column of Figure~\ref{fig2}(b). 
%The plasma bubble is not easily identifiable, but its front can be inferred from the background loops, as the magnetic flux rope rises, pushing aside and compressing the surrounding background loops.
%\textbf{[need to draw in fig2,3]This bubble is characterized by a brighter outer edge and a darker central cavity, and visible across most AIA channels.} % suggests its broad temperature range, spanning from 0.5 MK to over 10 MK. This broad spectral response may imply significant plasma heating and dynamic processes occurring within the structure.
%\textbf{Need a couple of sentences here to establish that the filament, cavity, and ``plasma bubble'' are the manifestation of an erupting magnetic flux rope.}
%The synchronous brightening of the main flare region and remote ribbons alongside the ascending plasma bubble may indicate a common origin within the same magnetic flux rope. 
According to the nearly east-west orientation of the flare ribbons shown in Figure~\ref{fig2}(c), the flux rope, located near the eastern solar limb, may present a view along its axis. %and appears to have an angle of inclination, consistent with three-dimensional flare models \citep{2022NatAs...6..317S}. 
%The erupting plasma bubble demonstrates rapid expansion in the north-south direction, which is perpendicular to the direction of eruption and aligns with the phase of impulsive acceleration. 

%%%%%bubble out of AIA view and flare ribbons
%At the final stage of the event, post-flare loops are clearly visible in the 211 \AA images, and their footpoints extend into two V-like shape ribbons or parallel ribbons in the main flare region in the 1600 \AA channel (see the right column in Figures~\ref{fig1}(a) and (b)). Such an arcade-ribbon configuration is consistent with the standard flare model. 

To further study the high-energy aspects of the erupting-MFR-driven event, we utilize X-ray imaging provided by the STIX instrument onboard Solar Orbiter. 
The SolO is located on the eastern side of the Sun, at an angular separation of approximately 18 degrees from the Sun–Earth line, with a heliocentric distance of about 0.95 AU.
The STIX images were generated using the STIX Ground Software\footnote{https://github.com/i4Ds/STIX-GSW}, with the \texttt{CLEAN} method utilized for image reconstruction.%\texttt{mem\_ge}
The X-ray imaging initially reveals a compact, semi-elliptical source at the onset of the flare, with an east–west extension in the 30\% contours of the nonthermal emission aligning with the "hook" region of the UV flare ribbons during the first microwave/HXR peak. However, due to limited spatial extent, temporal persistence, and limited dynamic range of STIX, we are unable to confirm this feature as a counterpart to the MFR footpoints, as reported by \citet{2023A&A...670A..89S}.
In Figure~\ref{fig2} (c4), AIA 1600 \AA\ images are projected and aligned to match the viewing coordinate of STIX. At the end of the main microwave/HXR peak (around 21:49:15 UT), the nonthermal emission shows the standard flare picture: the 32--76 keV nonthermal emission shows two footpoints that correspond to the conjugate UV ribbons, likely due to the precipitated nonthermal electrons, while the thermal emission (15--18 keV) shows the hot flare loop connecting the two footpoints, similar to \cite{2024SoPh..299..114R}.

\subsubsection{Microwave imaging spectroscopy}
\label{sec-obs-mw}

To derive microwave images from EOVSA datasets, several calibration steps are applied to the raw visibility data using well-developed routines. Initially, delay calibration, bandpass calibration, and gain calibration are performed. To enhance image quality, self-calibration techniques are employed. A circular beam with a full-width at half-maximum (FWHM) size of $90''/\nu$ GHz is used to restore the synthesized images. 
%\textbf{Flux scaling.}%Additionally, total power calibration is conducted at each frequency. This involves calculating a scaling factor by comparing the integrated flux from the imaging with the total power flux at peak times, which is then consistently applied across all time sequences.

Microwave spectroscopy imaging is performed at 33 equally-spaced frequencies ranging from 2.9 to 13.3 GHz with a temporal resolution of 4 seconds using EOVSA data. 
Initially, during the time of the HXR precursor when the filament begins its gradual ascent, compact microwave sources are observed near the activated filament region (Figure~\ref{fig3}(a)), which may arise from plasma heating or initially accelerated particles due to the associated energy release process.
%at the flare core, aligning with the configuration of the flare arcade (Figure~\ref{fig3}(a)).
These sources may trace initially accelerated energetic electrons at the flare site. %transport from the reconnection site, moving upwards into the ascending flux rope and downward towards the footpoints of the flare arcade.
During this phase, the plasma bubble is noted to remain at a relatively low altitude, making it hard to distinguish from the flare core.

Later on, the filament and plasma bubble ascend rapidly. Concurrently, the microwave sources begin to evolve with %elongate in both the northern-southern and eastern-western directions, aligning with the extension direction of 
the rising filament and moving synchronously with it, as indicated in Figures~\ref{fig3}(b)-(e). At later times, the microwave source nearly shapes the leg of the plasma bubble %signifying intense energy release and particle acceleration. The rising microwave sources follow the expansion of the flux rope and 
and displays a concave shape. This morphology may suggest that the microwave-emitting energetic electrons may have been injected into the MFR, possibly from the acceleration site below the MFR, and get trapped there.  %are possibly escaping, attempting to converge with the bubble and the flux rope along the reconnecting magnetic field lines. 
%\textbf{Is this also around the same time the HXR footpoint sources show an extension?} 
In the later stages, after around 21:44 UT, as the plasma bubble moves further away, the microwave emission returns to the main flare region (Figure~\ref{fig3}(f)), indicating that the accelerated electrons are again concentrated at the central flare site. 
%\textbf{[fig3? or another cartoon fig?]need a clear example to show the correspondence of the microwave source to the filament, the plasma bubble, and the cavity clearly, perhaps as a separate figure.}

%The variation in source positions at the peak intensity over time depicts the trajectory of the source. Initially, the sources are centered at the loop-top of the flare arcade at a relatively low altitude. As the filament eruption progresses, the sources ascend, followed by a descent as the flux rope moves away. In the post-flare phase, sources slightly rise with the thermalized post-flare arcade and stay stable at a higher position than their initial position. 
%are indicated Figure \ref{5eo_imgpk}, where clearly 

%%%%%GSfit
In order to constrain the microwave source properties, we conducted spatially resolved spectral analysis using brightness temperature spectra derived from three distinct regions of interest. One region is located near the filament core during its initiation phase (red box in Figure~\ref{fig3}(a)), and two others are located at the northern and southern legs of the flux rope during its eruption phase (blue and green boxes in Figure~\ref{fig3}(c), respectively).   
%(centered at the coordinate [-968, 80] with a width of 10 arcseconds), the southern (centered at [-978, 68] with a width of 10 arcseconds) and northern (centered at [-978, 100] with a width of 10 arcseconds) legs of the erupting bubble, marked as the red, green and blue boxes in Figure \ref{fig3} (a)(c). This analysis was aligned with two time intervals in the evolution of the source morphology: the single source during the initial flare trigger ($t_1$ at 21:38:48), the configuration shaping the bottom of the erupting bubble($t_2$ at 21:40:24).
% and the reversion to a single source after the bubble erupted away ($t_3$ at 21:43:24).
Spatially resolved microwave brightness temperature spectra derived from all three regions exhibited both positive and negative slopes with peaks at around 6--8 GHz, consistent with non-thermal gyrosynchrotron emission. 
To quantify the magnetic field strength ($B$), spectral index ($\delta'$), and the total number density ($n_e$) of non-thermal electrons, we employed the \texttt{pyGSFIT}\footnote{https://github.com/suncasa/pygsfit} tool to carry out the spectral fitting. Minimization is based on \texttt{SciPy}'s \texttt{optimize} package, and the gyrosynchrotron radiation calculation utilizes the fast gyrosynchrotron codes developed by \cite{2010ApJ...721.1127F}. The spectral fitting assumed a homogeneous source with a power-law electron energy distribution. The free fit parameters included the magnetic field strength $B$, non-thermal electron number density $n_e$, power-law index $\delta$, low-energy cutoff of the non-thermal electron distribution $E_{\rm min}$, and thermal plasma number density ($n_{\rm th}$). 

We then applied the Markov chain Monte Carlo (MCMC) method to robustly estimate the uncertainties of the fit parameters, following the approach outlined by \cite{2020NatAs...4.1140C}. The uncertainties of the measured brightness temperature spectra were considered and estimated by combining the root mean square (RMS) noise level in a selected region of the microwave images where no source is present, along with an assumed systematic error of 20\% in the absolute brightness temperature values (expected from the absolute flux calibration). The best-fit magnetic field strength, power-law index, and non-thermal electron number density are indicated on the top in Figures~\ref{fig3}(g)-(i).

%The magnetic field strength varies significantly from the flare core to the MFR legs, ranging from $408\pm74$ G at Box1 at $t_1$ to $339\pm53$ G and $296\pm41$G at Box2 and Box3 during $t_2$, respectively. 
The best-fit magnetic field strength of the erupting MFR near its legs low in the corona is around 300 G, consistent with a previous report by \citet{2020ApJ...895L..50C}. %This variation suggests a dispersion or weakening of the magnetic field as the flare progresses across different regions.
The power-law index of the nonthermal electron distribution hardens from $\sim$8.7 during the filament initiation phase at 21:38:48 UT to $\sim$6.0 during its erupting phase at 21:40:24 UT close to the first microwave/HXR peak, 
%to a softened $4.9\pm0.5$, $5.0\pm0.4$ at Box2, Box3 during $t_2$, suggesting an increasing proportion of higher energy electrons relative to lower energy electrons at $t_2$ as the flare evolves, 
consistent with the expectation of a more efficient electron acceleration process when the flare energy release intensifies \citep{2004A&A...426.1093G, 2011SSRv..159..107H}.
%that showing spectral soft-hard-soft evolution in rise-peak-decay phases
The similarity in the derived magnetic field strengths from the two regions of interest near the conjugate flux rope legs reaffirms that they may belong to the same flux rope. In addition, the similar values of the spectral indices of the source nonthermal electron distribution suggest a common origin of the source energetic electrons, which, as suggested by \citet{2020ApJ...895L..50C}, may be due to the transport of the energetic electrons, originated in the core flaring region, following along the flux rope field lines to its legs.%  indicate similar physical processes or similar electron acceleration mechanisms active across the lower and upper bottoms of the erupting bubble.}
We also note that \citet{2020ApJ...895L..50C} reported a harder nonthermal electron spectral index (2.5--2.7) in the MFR legs, whereas we find much softer spectra in this event. This difference likely reflects variations in acceleration efficiency, electron trapping, and event-specific factors such as flare properties, viewing geometry, and magnetic configuration.

%\textbf{The non-thermal electron number density varies from $10^{9.1\pm1.7} \text{cm}^{-3}$ at Box1 at $t_1$, to lower values in two boxes at higher positions at $t_2$, $10^{7.8\pm1.0} \text{ cm}^{-3}$ at Box2 and $10^{8.8\pm0.8} \text{ cm}^{-3}$ at Box3. This distribution suggests a reduction in the number of accelerated electrons and a broadening of the electron distribution region over time, indicating a bifurcation in the motion of electrons from the flare core to two distinct regions of the erupting bubble and more non-thermal electrons in northern legs of the erupting bubble (due to more close to the flare arcade? some electrons still remain in the main flare region).}

%%%%%%%%%%%%%%%%%%%%%%%%%%%%%%%%%%%%%%%%%%%%%%%%%%
\subsection{The Eruption in the Middle Corona}
\label{sec-obs-cme}
\subsubsection{White Light Observations}
The erupting bubble moves out of the field of view of the SDO/AIA at 21:45 UT (1.28$R_{\odot}$) but can still be viewed by GOES-R/SUVI at 21:47 UT, which has a larger field of view (around 1.67$R_{\odot}$). At 22:00 UT, the eruption appears in the LASCO/C2 field of view as a white light CME. %The main flare region is identified at coordinates [-960, 77] arcsec and is marked with a red diamond symbol in Figure \ref{fig4}, situated centrally between the two footpoints of the erupting bubble.
In order to obtain the trajectory of the erupting bubble, we tracked the leading edges using data from SDO/AIA, GOES/SUVI 131 \AA, and LASCO C2, C3 images at various times, indicated by dots from Figure~\ref{fig4}(a). Spline interpolation was employed to generate smooth trajectories of the bubble's front. The most distant point of the erupting bubble from the flare center at each time is denoted by a plus symbol.

The distance of the bubble's leading edge in time is shown in Figure~\ref{fig4}(b). 
The speed of the moving leading edge is estimated by using the difference in its position over time, as indicated by the blue cross symbols on the secondary y-axis. We also apply a linear fitting approach to get the speed of the bubble's front separately from the AIA, SUVI, and LASCO views. The results indicate that the speed was 242 km/s over a range of 0.02 to 0.13 $R_{\odot}$ from 21:38:42 to 21:44:18 as observed in the AIA 131 \AA view. Subsequently, a significantly higher speed of 1025 km/s was recorded between 0.14 and 0.50 $R_{\odot}$ from 21:43:49 to 21:47:49 in the SUVI 131 \AA\'s view. Lastly, the speed reached its peak at around 2807 km/s as the bubble expanded from 3.95 to 11.40 $R_{\odot}$ between 22:00:05 and 22:30:05, as captured in the LASCO view, which is consistent with the speed of 2852 km/s reported by the \textit{SOHO LASCO CME CATALOG}\footref{cme_catalog}.
%\footnote{\url{https://cdaw.gsfc.nasa.gov/CME_list/}}
These observations suggest a considerable acceleration of the bubble as it moves outward from the flare site and traverses the middle corona \citep{2001ApJ...559..452Z}. %The increasing velocities may imply dynamic interactions and potentially powerful driving mechanisms at play in the upper corona.

%The erupting bubbles erupt out and form a halo CME, which is defined to be surround the occulting disk of the coronagraph \citep{2010SunGe...5....7G}. It exhibits the classic three-part structure in the LASCO C2 white-light coronagraph images: a leading front, a dark enclosed cavity, and an embedded bright core. The evolution of the CME shows in three distinct phases: a slow rise phase, an impulsive acceleration phase, and a propagation phase at a near-constant velocity. These phases correspond to the three stages of the associated flare: pre-flare, rise, and decay phases, as detailed in the study by \cite{2001ApJ...559..452Z}.

%\subsection{GCS modelling}
%\label{sec-obs-GCS}

We employ the Graduated Cylindrical Shell (GCS) model \citep{2006ApJ...652..763T, 2011ApJS..194...33T} to reconstruct the magnetic structure of observed white light CME. The GCS model characterizes the CME as a combination of a tubular main body and a croissant-like 3D shape with conical legs. The geometry of this model, which assumes self-similar expansion, can be fully defined by three parameters: the angular half width $\alpha$, the aspect ratio $\kappa$, and the apex height $h_{apex}$.
We extracted the geometrical expressions from the Python code developed by \cite{forstner_2024_12668802}. The best-fit model for the CME observed at 22:00 UT is shown in Figure~\ref{fig4}(c) with $\alpha=50^{\circ}$, $\kappa=0.4$, $h_{apex}=3.92 R_{\odot}$.
%The parameters that yielded the best fit for the CME observed at 22:00 UT using LASCO data are as follows:
%the angular half width $\alpha=50^{\circ}$, 
%the aspect ratio $\kappa=0.4$, 
%the apex height $h_{apex}=3.92 R_{\odot}$,
%the rotation parameters used to align the CME with the observed positions are 
%the heliographic latitude $\theta=-15^{\circ}$ and stonyhurst longitude $\phi=-90^{\circ}$, and the tilt angle $\gamma=90^{\circ}$.
%The reconstructed meshed GCS model is projected onto the helioprojective coordinates, as shown in Figure \ref{fig4} (c).
The cross-section of the CME flux rope is a circular annulus with a varying radius $r_{c}(l)=\frac{\kappa}{\sqrt{1-\kappa^2}} l$, where $l$ is the length of the central axis for the conical legs. For the tubular part, the cross-section radius can be determined using Equations 18 and 19 in \citet{2011ApJS..194...33T}.

This best-fit GCS model of the CME is then extrapolated in a self-similar fashion to represent the location and morphology of the CME at times when LASCO data are not available.
Specifically, we keep all other model parameters constant but only vary $h_{apex}$ over time using the fitted trajectory of the leading edge of the MFR/CME, as shown in Figure~\ref{fig4}(b).
%We keep all other model parameters constant but only varying $h_{apex}$ over time using a propagation speed of 2807 km/s based on the observed white light CME speed.
Figure~\ref{fig4}(d) shows an example of the extrapolated CME model at an earlier time (21:52:36 UT) overlayed on LASCO/C2 image at 21:48:07 UT, when the CME cannot be identified directly from LASCO data as it remains under the coronagraph's occulter.
%when the CME has not entered LASCO's field of view yet
This model at 21:52:36 UT will be used to compare with the meterwave spectral imaging data obtained by OVRO-LWA, discussed next.  %The apex height can be determined from the CME's propagation with a constant speed.
%We developed a function to calculate the cross-section radius at any given position within the GCS shell. For a specified location, the function identifies the nearest cross-section plane and returns the corresponding radius. 

%The metric-wave imaging at 21:52:36 UT will be further analyzed and applied spectral fitting. The morphology of the GCS model at this time, with an apex height of $h_{apex}=1.83 R_{\odot}$, is presented in Figure \ref{fig4}(d). The input point, located at [-2535, -348] arcsec, corresponds to the center of Box1, which is selected for spectral fitting of metric-wave imaging. The cross-sectional radius at this input point is $0.52 R_{\odot}$, as marked by a green circle in Figure \ref{fig4}(d).

%%%%%%%%%%%%%%%%%%%%%%%%%%%%%%%%%%%%%%%%%%%%%%%%%%
\subsubsection{Meter-wavelength imaging spectroscopy}
\label{sec-obs-radio}

%\textbf{Imaging data intro} 

Meterwave spectral imaging with OVRO-LWA data was performed at seven frequency bands of 32, 41, 46, 55, 69, 78, and 82 MHz with a time resolution of 10 seconds from 21:41:00 to 22:10:00 UT using the solar data pipeline (Mondal et al., in preparation). Each image was made by integrating over the respective frequency band with a bandwidth of \~{}5 MHz. The frequency-dependent synthesized beam sizes are shown as eclipses in Figure~\ref{fig5}. At the highest frequency (82 MHz), the FWMH along the major and minor axes of the beam at 21:41:00 UT is $12'$ and $5'$, respectively.  
%Minor and major beam sizes and inclination angle changes at 32 MHz ranged from 10.23, 20.61 arcmins and -71.27 degrees, to 10.10,  23.13 arcmins and -63.66 degrees, respectively.  At 82 MHz, these changed from 4.81, 11.77 arcmins and -69.94 degrees, to 4.72, 14.00 arcmins and -63.33 degrees.
%Dynamic spectral data from the beamforming mode was recorded across a frequency range of 22-86 MHz with a resolution of 23.925 kHz and a temporal resolution of 65 ms within the same time frame.

%\textbf{Refrection} 
Due to the influence of ionospheric refraction effects, the apparent source position from OVRO-LWA imaging is shifted following a $1/\nu^2$ frequency dependence \citep[e.g.,][]{2009_cohen_AJ, loi_2015_GeoRL, helmholdt_2020_RaSc}. To correct for these effects, we fitted the solar disk using a Gaussian function to determine the centroid shifts at each frequency band during a quiescent time just prior to the event (21:41:14 UT, shown in Figure~\ref{fig5}(a)). Notably, there is a small off-limb narrow-band burst seen at $<$60 MHz at this time, which is excluded from the disk fitting.
These shifts were then applied across the whole time interval. 
Even though our analysis revealed only a very small variation in the solar center position before and after the burst, with the two appearing nearly/visually identical and well within the average beam size of 10 arcminutes, a refraction-induced shift may still occur during the two quiet Sun intervals. As such, our measurements should be considered a lower limit under this approach.

As shown in Figure~\ref{fig1}(d), OVRO-LWA has observed a rich variety of complex metric bursts during this event. Most of the bursts are coherent, and they may be the topic of future investigations. In the present work, we selected a time at 21:52:36 during which no significant coherent radio bursts are evident in the dynamic spectrum. 
The multi-frequency metric-decametric radio source at this time generally aligns with the extrapolated GCS model of the white-light CME, as illustrated in Figure~\ref{fig5}, although there is some extension beyond the CME model, particularly toward the southern part of the CME at low frequencies. By examining images made at nearby times when bright coherent bursts are present, we attribute the extension to relatively weak coherent emission near the front of the CME, possibly associated with a CME-driven shock.

To investigate the spectral properties of the metric sources, we selected several regions of interest, each with a size of 8 arcmin, positioned to sample the CME front (blue, orange, green boxes; regions 1-3), CME core (red box; region 4), and flare site (purple box; region 5). The brightness temperature within these boxes was averaged, and the resulting light curves are displayed in Figures~\ref{fig5}(f)-(h).
Our analysis reveals that gyrosynchrotron-like spectra predominantly appear at the CME front, specifically within regions 1 and 3. The observed spectrum displays a peak at approximately 50 MHz and a maximum brightness temperature around 600 MK, indicative of gyrosynchrotron emission. 
The best-fit parameters for the brightness temperature spectrum from region 1 are  
magnetic field strength \( B = 0.6 \pm 0.1 \, \text{G} \), 
non-thermal electron number density \( n_e^{\text{nth}} = 10^{4.7} \pm 10^{1.0} \, \text{cm}^{-3} \), and power-law index \( \delta = 8.7 \pm 0.6 \). We will discuss the implications of these results in conjunction with the microwave diagnostics of the eruption in the low corona in the next section.

\begin{figure*}%[!ht]
\centering
\includegraphics[width=15cm]{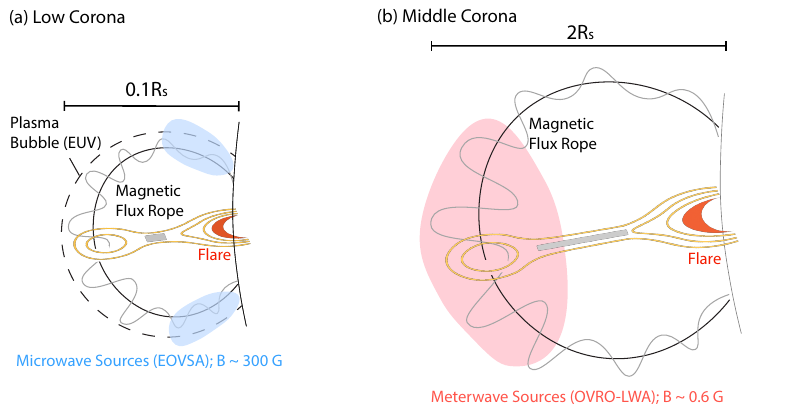}
\caption{Schematic diagram of (a) the microwave sources associated with the MFR during its initiation phase in the lower corona, and (b) metric-wave sources related to the MFR as it propagates into the middle corona.
While the cartoon is approximately scaled in relative spatial extent, the actual heights of the CME leading edge in panels (a) and (b) are about 0.1 $R_{\odot}$ and 2.0 $R_{\odot}$, respectively.
\label{fig6}}
\end{figure*}

%%%%%%%%%%%%%%%%%%%%%%%%%%%%%%%%%%%%%%%%%%%%%%%%%%

\section{Summary and Discussions}
\label{sec-dis}

In this paper, we present multi-wavelength observations of a CME event associated with an X5-class flare from the low to middle corona.
%a comprehensive multiwavelength analysis of an X5-class solar flare that follows the standard flare model, featuring a filament eruption associated with a magnetic flux rope. 
In particular, microwave and radio imaging spectroscopy observations reveal counterparts of the erupting-MFR-driven CME, as indicated from Figures~\ref{fig6}(a) and (b). The emission mechanism is identified as gyrosynchrotron radiation produced by accelerated electrons gyrating around the magnetic field lines that comprise the MFR, offering a unique diagnostic tool for diagnosing the magnetic field from the low to middle corona. %This study provides, for the first time, the measurement of the magnetic field strength of the magnetic flux rope using both microwave and radio imaging spectroscopy.

%The multiwavelength observations significantly enhance our understanding of the eruptive flare processes. 
During the initiation phase of the event, a filament is activated and rises upwards. It represents the lower portion of a rising MFR. This ascent stretches the magnetic field lines, triggering magnetic reconnection, which then facilitates the energy release, leading to the X-class flare.
Microwave spectral imaging obtained by EOVSA reveals that the source during the early ascending phase of the MFR displays a concave shape that outlines the bottom of the MFR cavity. %As the erupting bubble moves away, the MW emission return to focus above the post-flare core. 
Similar to \citet{2020ApJ...895L..50C}, we interpret the peculiar appearance of the microwave source as flare-accelerated nonthermal electrons trapped near the MFR legs. This picture is further supported by the similarity of the derived magnetic field strength ($\sim$300 G) and nonthermal electron properties for the northern and southern parts of the microwave source.
OVRO-LWA captured a variety of bright metric radio bursts during the event. During a relatively quiescent period, we find that the radio source closely follows the extrapolated CME morphology in the middle corona, despite the fact that additional radio sources are still present, rendering the observed radio source to extend slightly beyond the extrapolated CME body.
%\textcolor{red}{In Figure \ref{fig5}(b) and (e), the upper/front edges trace the CME’s leading front, while the lower portion corresponds to rooted loops, with the current sheet situated in between. We note that radio imaging does not precisely match the sizes seen in optical observations, such as the quiet Sun, and often appears larger due to the combined effects of contributions from the CME’s leading edges, potential projection and scattering effects, and the lower spatial resolution at lower frequencies.}
Spectral analysis indicates that the source near the CME front is best described as gyrosynchrotron radiation from highly energetic, hundreds of keV electrons. The exact origin of these electrons is debatable, which may either be injected into the MFR body from the flare site \citep{2001ApJ...558L..65B}, or accelerated by the CME-driven shock but convected to the shock downstream \citep{2022A&A...668A..15M}. In either case, the presence of these energetic electrons gives rise to bright gyrosynchrotron radiation, allowing us to constrain the magnetic field strength in the MFR body to be around 0.6 G at a radial distance of 1.8 \( R_{\odot} \).

%%==========Magnetic strength evolution
%From metricwave imaging spectral fitting, the magnetic field strength is approximately 2 G at a projected height of 2.83 \( R_{\odot} \), while Box1 is located at \([-2535, -348]\) arcsecs, with a cross-sectional radius of 0.52 \( R_{\odot} \) (equivalent to 500 arcsec), as shown in Figure \ref{fig4}(d). 
Can we reconcile the magnetic field measurements of the CME from the low to middle corona?
Assuming the magnetic flux of the MFR, defined as $\Phi = B^{t}A$, where $B^{t}$ represents the toroidal component of the MFR's magnetic field, and $A$ denotes the cross-section area of the MFR, is conserved, we have the following relation:

\begin{equation}
\label{eq:B_flux}
% \frac{B^{t}_{\rm low}}{B^{t}_{\rm mid}} = \frac{A_{\rm mid}}{A_{\rm{low}}},
B^{t}_{\rm low}A_{\rm{low}}=B^{t}_{\rm mid}A_{\rm mid},
\end{equation}

where $B^{t}_{\rm low}\approx 300$~G and $A_{\rm low}\approx \pi R_{\rm low}^2$ represent the magnetic field strength and cross-sectional area at the MFR leg during its initiation phase in the low corona when the microwave sources are observed.
In the middle corona, the cross-sectional radius is $\sim$0.52 $R_{\odot}$ for region 1 and $\sim$0.50 $R_{\odot}$ for region 3, with corresponding magnetic field strengths of $\sim$0.6 G and $\sim$0.7 G, respectively, the estimated magnetic fluxes are comparable. We therefore use region 1 as a representative case for estimating magnetic flux in the middle corona.
$B^{t}_{\rm mid}\approx 0.6$~G and $A_{\rm mid}\approx \pi R_{\rm mid}^2$ are the corresponding values when the CME propagates into the middle corona being observed in meterwaves. %Solving for \( R_{\text{low}} \) yields approximately 30 arcsec for the legs of the MFR at a projected height of 1.02 \( R_{\odot} \), which is reasonable from EUV observations.
The GCS model, which assumes conical legs originating from the flare site with the apex projected as a single point on the solar surface, does not fully capture the realistic morphology of MFR leg footpoints while the filament is close to the flare site. However, the GCS model may be adequately used as an estimate for the MFR cross-section $A_{\rm mid}$ in the middle corona when the CME is well-developed. The best-fit GCS model at the time of the meterwave source suggests $A_{\rm mid} \approx 4.1\times10^{21}\ \mathrm{cm}^2$ at the apex of the MFR. Solving for $A_{\text{low}}$ in Equation~\ref{eq:B_flux} yields $R_{\rm low}\approx 1.6\times10^9$ cm, or $\sim\!22''$, which corresponds to the cross-section size of the MFR leg in the low corona where microwave diagnostics are available. To corroborate with our estimate, we draw a slice for each of the two MFR legs perpendicular to the presumed MFR spine (the dashed lines in Figure \ref{fig3}(c)) and fitted the flux profile with a Gaussian function to determine the full width at half maximum (FWHM), from which the cross-sectional radius is taken as half of the FWHM.
At 5.8 GHz, the FWHM is approximately 40$''$, or $\sim\,2R_{\rm low}$, closely aligning with theoretical expectations assuming magnetic flux conservation.
%The leading edge of MFR farthest from the flare center is approximately at [−1050,75] inferred from EUV observations. The cross-sectional radius of the microwave source at the MFR legs at this height is estimated to be around 10 arcseconds.
%and \theta is the angle between the magnetic field lines and the normal (perpendicular) to S.
%These analyses of magnetic field strength gradients and MFR cross-section may offer direct evidence supporting the near-conservation of magnetic flux, 
Our measured magnetic field strength of the evolving CME from low to middle corona that significantly follows the law of magnetic flux conservation offers direct support for the MFR interpretation of CMEs, and highlights the capability of combined microwave and radio imaging spectroscopy to diagnose magnetic fields of CMEs from its birth to its propagation into the middle corona.

%Building on these findings, the spatially resolved measurements of the magnetic field strength offer critical insights into the evolution of MFRs and their role in driving solar eruptions. With the expectation of next-generation solar radio observations and advanced modeling techniques, the integration of multi-wavelength diagnostics will further refine our understanding of the Sun's magnetic processes. Such approaches are essential for advancing space weather prediction capabilities and unveiling the complex dynamics of the heliosphere.

%%%%%%%%%%%%%%%%%%%%%%%%%%%%%%%%%%%%%%%%%%%%%%%%%%
\begin{acknowledgements}
X.C., B.C., S.Y., S.M., and D.G. were supported by NASA grants 80NSSC20K0026 and 80NSSC24K1116 to the New Jersey Institute of Technology (NJIT). EOVSA was designed, built, and is now operated by NJIT as a community facility. The EOVSA operations are supported by NSF grants AGS-2130832 and NASA grant 80NSSC20K0026 to NJIT. The OVRO-LWA expansion project was supported by NSF under grant AST-1828784. OVRO-LWA operations for solar and space weather sciences are supported by NSF under grant AGS-2436999.
The Solar Orbiter is a space mission of international collaboration between ESA and NASA, operated by ESA. The STIX instrument is an international collaboration between Switzerland, Poland, France, the Czech Republic, Germany, Austria, Ireland, and Italy.
This CME catalog is generated and maintained at the CDAW Data Center by NASA and The Catholic University of America in cooperation with the Naval Research Laboratory. SOHO is a project of international cooperation between ESA and NASA.
\end{acknowledgements}

%%%%%%%%%%%%%%%%%%%%%%%%%%%%%%%%%%%%%%%%%%%%%%%%%%

\bibliography{xflare_MFR}{}

\begin{thebibliography}{}
\expandafter\ifx\csname natexlab\endcsname\relax\def\natexlab#1{#1}\fi
\providecommand{\url}[1]{\href{#1}{#1}}
\providecommand{\dodoi}[1]{doi:~\href{http://doi.org/#1}{\nolinkurl{#1}}}
\providecommand{\doeprint}[1]{\href{http://ascl.net/#1}{\nolinkurl{http://ascl.net/#1}}}
\providecommand{\doarXiv}[1]{\href{https://arxiv.org/abs/#1}{\nolinkurl{https://arxiv.org/abs/#1}}}

\bibitem[{M.~M. {Anderson} {et~al.}(2018){Anderson}, {Hallinan}, {Eastwood},
  {Monroe}, {Vedantham}, {Bourke}, {Greenhill}, {Kocz}, {Lazio}, {Price},
  {Schinzel}, {Wang}, \& {Woody}}]{2018ApJ...864...22A}
{Anderson}, M.~M., {Hallinan}, G., {Eastwood}, M.~W., {et~al.} 2018,
  \bibinfo{title}{{A Simultaneous Search for Prompt Radio Emission Associated
  with the Short GRB 170112A Using the All-sky Imaging Capability of the
  OVRO-LWA},} \apj, 864, 22, \dodoi{10.3847/1538-4357/aad2d7}

\bibitem[{H.~M. {Bain} {et~al.}(2014){Bain}, {Krucker}, {Saint-Hilaire}, \&
  {Raftery}}]{2014ApJ...782...43B}
{Bain}, H.~M., {Krucker}, S., {Saint-Hilaire}, P., \& {Raftery}, C.~L. 2014,
  \bibinfo{title}{{Radio Imaging of a Type IVM Radio Burst on the 14th of
  August 2010},} \apj, 782, 43, \dodoi{10.1088/0004-637X/782/1/43}

\bibitem[{T.~S. {Bastian} {et~al.}(1998){Bastian}, {Benz}, \&
  {Gary}}]{1998ARA&A..36..131B}
{Bastian}, T.~S., {Benz}, A.~O., \& {Gary}, D.~E. 1998, \bibinfo{title}{{Radio
  Emission from Solar Flares},} \araa, 36, 131,
  \dodoi{10.1146/annurev.astro.36.1.131}

\bibitem[{T.~S. {Bastian} {et~al.}(2001){Bastian}, {Pick}, {Kerdraon}, {Maia},
  \& {Vourlidas}}]{2001ApJ...558L..65B}
{Bastian}, T.~S., {Pick}, M., {Kerdraon}, A., {Maia}, D., \& {Vourlidas}, A.
  2001, \bibinfo{title}{{The Coronal Mass Ejection of 1998 April 20: Direct
  Imaging at Radio Wavelengths},} \apjl, 558, L65, \dodoi{10.1086/323421}

\bibitem[{J.~L. {Bougeret} {et~al.}(2008){Bougeret}, {Goetz}, {Kaiser}, {Bale},
  {Kellogg}, {Maksimovic}, {Monge}, {Monson}, {Astier}, {Davy}, {Dekkali},
  {Hinze}, {Manning}, {Aguilar-Rodriguez}, {Bonnin}, {Briand}, {Cairns},
  {Cattell}, {Cecconi}, {Eastwood}, {Ergun}, {Fainberg}, {Hoang}, {Huttunen},
  {Krucker}, {Lecacheux}, {MacDowall}, {Macher}, {Mangeney}, {Meetre},
  {Moussas}, {Nguyen}, {Oswald}, {Pulupa}, {Reiner}, {Robinson}, {Rucker},
  {Salem}, {Santolik}, {Silvis}, {Ullrich}, {Zarka}, \&
  {Zouganelis}}]{2008SSRv..136..487B}
{Bougeret}, J.~L., {Goetz}, K., {Kaiser}, M.~L., {et~al.} 2008,
  \bibinfo{title}{{S/WAVES: The Radio and Plasma Wave Investigation on the
  STEREO Mission},} \ssr, 136, 487, \dodoi{10.1007/s11214-007-9298-8}

\bibitem[{G.~E. {Brueckner} {et~al.}(1995){Brueckner}, {Howard}, {Koomen},
  {Korendyke}, {Michels}, {Moses}, {Socker}, {Dere}, {Lamy}, {Llebaria},
  {Bout}, {Schwenn}, {Simnett}, {Bedford}, \& {Eyles}}]{1995SoPh..162..357B}
{Brueckner}, G.~E., {Howard}, R.~A., {Koomen}, M.~J., {et~al.} 1995,
  \bibinfo{title}{{The Large Angle Spectroscopic Coronagraph (LASCO)},}
  \solphys, 162, 357, \dodoi{10.1007/BF00733434}

\bibitem[{E.~P. {Carley} {et~al.}(2017){Carley}, {Vilmer}, {Sim{\~o}es}, \&
  {{\'O} Fearraigh}}]{2017A&A...608A.137C}
{Carley}, E.~P., {Vilmer}, N., {Sim{\~o}es}, P. J.~A., \& {{\'O} Fearraigh}, B.
  2017, \bibinfo{title}{{Estimation of a coronal mass ejection magnetic field
  strength using radio observations of gyrosynchrotron radiation},} \aap, 608,
  A137, \dodoi{10.1051/0004-6361/201731368}

\bibitem[{B. {Chen} {et~al.}(2020{\natexlab{a}}){Chen}, {Yu}, {Reeves}, \&
  {Gary}}]{2020ApJ...895L..50C}
{Chen}, B., {Yu}, S., {Reeves}, K.~K., \& {Gary}, D.~E. 2020{\natexlab{a}},
  \bibinfo{title}{{Microwave Spectral Imaging of an Erupting Magnetic Flux
  Rope: Implications for the Standard Solar Flare Model in Three Dimensions},}
  \apjl, 895, L50, \dodoi{10.3847/2041-8213/ab901a}

\bibitem[{B. {Chen} {et~al.}(2020{\natexlab{b}}){Chen}, {Shen}, {Gary},
  {Reeves}, {Fleishman}, {Yu}, {Guo}, {Krucker}, {Lin}, {Nita}, \&
  {Kong}}]{2020NatAs...4.1140C}
{Chen}, B., {Shen}, C., {Gary}, D.~E., {et~al.} 2020{\natexlab{b}},
  \bibinfo{title}{{Measurement of magnetic field and relativistic electrons
  along a solar flare current sheet},} Nature Astronomy, 4, 1140,
  \dodoi{10.1038/s41550-020-1147-7}

\bibitem[{P.~F. {Chen}(2011){Chen}}]{2011LRSP....8....1C}
{Chen}, P.~F. 2011, \bibinfo{title}{{Coronal Mass Ejections: Models and Their
  Observational Basis},} Living Reviews in Solar Physics, 8, 1,
  \dodoi{10.12942/lrsp-2011-1}

\bibitem[{P.~F. {Chen} \& K. {Shibata}(2000){Chen} \&
  {Shibata}}]{2000ApJ...545..524C}
{Chen}, P.~F., \& {Shibata}, K. 2000, \bibinfo{title}{{An Emerging Flux Trigger
  Mechanism for Coronal Mass Ejections},} \apj, 545, 524,
  \dodoi{10.1086/317803}

\bibitem[{X. {Cheng} {et~al.}(2017){Cheng}, {Guo}, \&
  {Ding}}]{2017ScChD..60.1383C}
{Cheng}, X., {Guo}, Y., \& {Ding}, M. 2017, \bibinfo{title}{{Origin and
  Structures of Solar Eruptions I: Magnetic Flux Rope},} Science China Earth
  Sciences, 60, 1383, \dodoi{10.1007/s11430-017-9074-6}

\bibitem[{S. {Chhabra} {et~al.}(2021){Chhabra}, {Gary}, {Hallinan}, {Anderson},
  {Chen}, {Greenhill}, \& {Price}}]{Chhabra2021}
{Chhabra}, S., {Gary}, D.~E., {Hallinan}, G., {et~al.} 2021,
  \bibinfo{title}{{Imaging Spectroscopy of CME-associated Solar Radio Bursts
  using OVRO-LWA},} \apj, 906, 132, \dodoi{10.3847/1538-4357/abc94b}

\bibitem[{A.~S. {Cohen} \& H.~J.~A. {R{\"o}ttgering}(2009){Cohen} \&
  {R{\"o}ttgering}}]{2009_cohen_AJ}
{Cohen}, A.~S., \& {R{\"o}ttgering}, H.~J.~A. 2009, \bibinfo{title}{{Probing
  Fine-Scale Ionospheric Structure with the Very Large Array Radio Telescope},}
  \aj, 138, 439, \dodoi{10.1088/0004-6256/138/2/439}

\bibitem[{G.~A. {Dulk}(1985){Dulk}}]{1985ARA&A..23..169D}
{Dulk}, G.~A. 1985, \bibinfo{title}{{Radio emission from the sun and stars.},}
  \araa, 23, 169, \dodoi{10.1146/annurev.aa.23.090185.001125}

\bibitem[{G.~D. {Fleishman} \& A.~A. {Kuznetsov}(2010){Fleishman} \&
  {Kuznetsov}}]{2010ApJ...721.1127F}
{Fleishman}, G.~D., \& {Kuznetsov}, A.~A. 2010, \bibinfo{title}{{Fast
  Gyrosynchrotron Codes},} \apj, 721, 1127,
  \dodoi{10.1088/0004-637X/721/2/1127}

\bibitem[{J.~L. Forstner(2024)Forstner}]{forstner_2024_12668802}
Forstner, J.~L. 2024, \bibinfo{title}{GCS in Python,}, 0.2.3 Zenodo,
  \dodoi{10.5281/zenodo.12668802}

\bibitem[{D.~E. {Gary}(2023){Gary}}]{2023ARA&A..61..427G}
{Gary}, D.~E. 2023, \bibinfo{title}{{New Insights from Imaging Spectroscopy of
  Solar Radio Emission},} \araa, 61, 427,
  \dodoi{10.1146/annurev-astro-071221-052744}

\bibitem[{D.~E. {Gary} {et~al.}(2018){Gary}, {Chen}, {Dennis}, {Fleishman},
  {Hurford}, {Krucker}, {McTiernan}, {Nita}, {Shih}, {White}, \&
  {Yu}}]{Gary2018}
{Gary}, D.~E., {Chen}, B., {Dennis}, B.~R., {et~al.} 2018,
  \bibinfo{title}{{Microwave and Hard X-Ray Observations of the 2017 September
  10 Solar Limb Flare},} \apj, 863, 83, \dodoi{10.3847/1538-4357/aad0ef}

\bibitem[{P.~C. {Grigis} \& A.~O. {Benz}(2004){Grigis} \&
  {Benz}}]{2004A&A...426.1093G}
{Grigis}, P.~C., \& {Benz}, A.~O. 2004, \bibinfo{title}{{The spectral evolution
  of impulsive solar X-ray flares},} \aap, 426, 1093,
  \dodoi{10.1051/0004-6361:20041367}

\bibitem[{J.~F. {Helmboldt} \& N. {Hurley-Walker}(2020){Helmboldt} \&
  {Hurley-Walker}}]{helmholdt_2020_RaSc}
{Helmboldt}, J.~F., \& {Hurley-Walker}, N. 2020, \bibinfo{title}{{Ionospheric
  Irregularities Observed During the GLEAM Survey},} Radio Science, 55, e07106,
  \dodoi{10.1029/2020RS007106}

\bibitem[{G.~D. {Holman} {et~al.}(2011){Holman}, {Aschwanden}, {Aurass},
  {Battaglia}, {Grigis}, {Kontar}, {Liu}, {Saint-Hilaire}, \&
  {Zharkova}}]{2011SSRv..159..107H}
{Holman}, G.~D., {Aschwanden}, M.~J., {Aurass}, H., {et~al.} 2011,
  \bibinfo{title}{{Implications of X-ray Observations for Electron Acceleration
  and Propagation in Solar Flares},} \ssr, 159, 107,
  \dodoi{10.1007/s11214-010-9680-9}

\bibitem[{Q. {Hu} {et~al.}(2014){Hu}, {Qiu}, {Dasgupta}, {Khare}, \&
  {Webb}}]{2014ApJ...793...53H}
{Hu}, Q., {Qiu}, J., {Dasgupta}, B., {Khare}, A., \& {Webb}, G.~M. 2014,
  \bibinfo{title}{{Structures of Interplanetary Magnetic Flux Ropes and
  Comparison with Their Solar Sources},} \apj, 793, 53,
  \dodoi{10.1088/0004-637X/793/1/53}

\bibitem[{C. {Jiang} {et~al.}(2022){Jiang}, {Feng}, {Guo}, \&
  {Hu}}]{2022Innov...300236J}
{Jiang}, C., {Feng}, X., {Guo}, Y., \& {Hu}, Q. 2022,
  \bibinfo{title}{{Data-driven modeling of solar coronal magnetic field
  evolution and eruptions},} The Innovation, 3, 100236,
  \dodoi{10.1016/j.xinn.2022.100236}

\bibitem[{C. {Jiang} {et~al.}(2014){Jiang}, {Wu}, {Feng}, \&
  {Hu}}]{2014ApJ...786L..16J}
{Jiang}, C., {Wu}, S.~T., {Feng}, X., \& {Hu}, Q. 2014,
  \bibinfo{title}{{Nonlinear Force-free Field Extrapolation of a Coronal
  Magnetic Flux Rope Supporting a Large-scale Solar Filament from a
  Photospheric Vector Magnetogram},} \apjl, 786, L16,
  \dodoi{10.1088/2041-8205/786/2/L16}

\bibitem[{D. {Kansabanik} {et~al.}(2023){Kansabanik}, {Mondal}, \&
  {Oberoi}}]{2023ApJ...950..164K}
{Kansabanik}, D., {Mondal}, S., \& {Oberoi}, D. 2023,
  \bibinfo{title}{{Deciphering Faint Gyrosynchrotron Emission from a Coronal
  Mass Ejection Using Spectropolarimetric Radio Imaging},} \apj, 950, 164,
  \dodoi{10.3847/1538-4357/acc385}

\bibitem[{D. {Kansabanik} {et~al.}(2024){Kansabanik}, {Mondal}, \&
  {Oberoi}}]{2024ApJ...968...55K}
{Kansabanik}, D., {Mondal}, S., \& {Oberoi}, D. 2024,
  \bibinfo{title}{{Spectropolarimetric Radio Imaging of Faint Gyrosynchrotron
  Emission from a CME: A Possible Indication of the Insufficiency of
  Homogeneous Models},} \apj, 968, 55, \dodoi{10.3847/1538-4357/ad43e9}

\bibitem[{S. {Krucker} {et~al.}(2020){Krucker}, {Hurford}, {Grimm}, {K{\"o}gl},
  {Gr{\"o}belbauer}, {Etesi}, {Casadei}, {Csillaghy}, {Benz}, {Arnold},
  {Molendini}, {Orleanski}, {Schori}, {Xiao}, {Kuhar}, {Hochmuth}, {Felix},
  {Schramka}, {Marcin}, {Kobler}, {Iseli}, {Dreier}, {Wiehl}, {Kleint},
  {Battaglia}, {Lastufka}, {Sathiapal}, {Lapadula}, {Bednarzik}, {Birrer},
  {Stutz}, {Wild}, {Marone}, {Skup}, {Cichocki}, {Ber}, {Rutkowski}, {Bujwan},
  {Juchnikowski}, {Winkler}, {Darmetko}, {Michalska}, {Seweryn}, {Bia{\l}ek},
  {Osica}, {Sylwester}, {Kowalinski}, {{\'S}cis{\l}owski}, {Siarkowski},
  {St{\k{e}}{\'s}licki}, {Mrozek}, {Podg{\'o}rski}, {Meuris}, {Limousin},
  {Gevin}, {Le Mer}, {Brun}, {Strugarek}, {Vilmer}, {Musset}, {Maksimovi{\'c}},
  {F{\'a}rn{\'\i}k}, {Koz{\'a}{\v{c}}ek}, {Ka{\v{s}}parov{\'a}}, {Mann},
  {{\"O}nel}, {Warmuth}, {Rendtel}, {Anderson}, {Bauer}, {Dionies}, {Paschke},
  {Pl{\"u}schke}, {Woche}, {Schuller}, {Veronig}, {Dickson}, {Gallagher},
  {Maloney}, {Bloomfield}, {Piana}, {Massone}, {Benvenuto}, {Massa},
  {Schwartz}, {Dennis}, {van Beek}, {Rodr{\'\i}guez-Pacheco}, \&
  {Lin}}]{2020A&A...642A..15K}
{Krucker}, S., {Hurford}, G.~J., {Grimm}, O., {et~al.} 2020,
  \bibinfo{title}{{The Spectrometer/Telescope for Imaging X-rays (STIX)},}
  \aap, 642, A15, \dodoi{10.1051/0004-6361/201937362}

\bibitem[{J.~R. {Lemen} {et~al.}(2012){Lemen}, {Title}, {Akin}, {Boerner},
  {Chou}, {Drake}, {Duncan}, {Edwards}, {Friedlaender}, {Heyman}, {Hurlburt},
  {Katz}, {Kushner}, {Levay}, {Lindgren}, {Mathur}, {McFeaters}, {Mitchell},
  {Rehse}, {Schrijver}, {Springer}, {Stern}, {Tarbell}, {Wuelser}, {Wolfson},
  {Yanari}, {Bookbinder}, {Cheimets}, {Caldwell}, {Deluca}, {Gates}, {Golub},
  {Park}, {Podgorski}, {Bush}, {Scherrer}, {Gummin}, {Smith}, {Auker},
  {Jerram}, {Pool}, {Soufli}, {Windt}, {Beardsley}, {Clapp}, {Lang}, \&
  {Waltham}}]{2012SoPh..275...17L}
{Lemen}, J.~R., {Title}, A.~M., {Akin}, D.~J., {et~al.} 2012,
  \bibinfo{title}{{The Atmospheric Imaging Assembly (AIA) on the Solar Dynamics
  Observatory (SDO)},} \solphys, 275, 17, \dodoi{10.1007/s11207-011-9776-8}

\bibitem[{R. {Liu}(2020){Liu}}]{2020RAA....20..165L}
{Liu}, R. 2020, \bibinfo{title}{{Magnetic flux ropes in the solar corona:
  structure and evolution toward eruption},} Research in Astronomy and
  Astrophysics, 20, 165, \dodoi{10.1088/1674-4527/20/10/165}

\bibitem[{S.~T. {Loi} {et~al.}(2015){Loi}, {Murphy}, {Cairns}, {Menk},
  {Waters}, {Erickson}, {Trott}, {Hurley-Walker}, {Morgan}, {Lenc}, {Offringa},
  {Bell}, {Ekers}, {Gaensler}, {Lonsdale}, {Feng}, {Hancock}, {Kaplan},
  {Bernardi}, {Bowman}, {Briggs}, {Cappallo}, {Deshpande}, {Greenhill},
  {Hazelton}, {Johnston-Hollitt}, {McWhirter}, {Mitchell}, {Morales}, {Morgan},
  {Oberoi}, {Ord}, {Prabu}, {Shankar}, {Srivani}, {Subrahmanyan}, {Tingay},
  {Wayth}, {Webster}, {Williams}, \& {Williams}}]{loi_2015_GeoRL}
{Loi}, S.~T., {Murphy}, T., {Cairns}, I.~H., {et~al.} 2015,
  \bibinfo{title}{{Real-time imaging of density ducts between the plasmasphere
  and ionosphere},} \grl, 42, 3707, \dodoi{10.1002/2015GL063699}

\bibitem[{D.~J.~F. {Maia} {et~al.}(2007){Maia}, {Gama}, {Mercier}, {Pick},
  {Kerdraon}, \& {Karlick{\'y}}}]{2007ApJ...660..874M}
{Maia}, D. J.~F., {Gama}, R., {Mercier}, C., {et~al.} 2007,
  \bibinfo{title}{{The Radio-Coronal Mass Ejection Event on 2001 April 15},}
  \apj, 660, 874, \dodoi{10.1086/508011}

\bibitem[{S. {Mondal} {et~al.}(2020){Mondal}, {Oberoi}, \&
  {Vourlidas}}]{2020ApJ...893...28M}
{Mondal}, S., {Oberoi}, D., \& {Vourlidas}, A. 2020,
  \bibinfo{title}{{Estimation of the Physical Parameters of a CME at High
  Coronal Heights Using Low-frequency Radio Observations},} \apj, 893, 28,
  \dodoi{10.3847/1538-4357/ab7fab}

\bibitem[{D.~E. {Morosan} {et~al.}(2022){Morosan}, {Pomoell}, {Kumari},
  {Vainio}, \& {Kilpua}}]{2022A&A...668A..15M}
{Morosan}, D.~E., {Pomoell}, J., {Kumari}, A., {Vainio}, R., \& {Kilpua},
  E.~K.~J. 2022, \bibinfo{title}{{Shock-accelerated electrons during the fast
  expansion of a coronal mass ejection},} \aap, 668, A15,
  \dodoi{10.1051/0004-6361/202244432}

\bibitem[{D.~F. {Ryan} {et~al.}(2024){Ryan}, {Massa}, {Battaglia}, {Dickson},
  {Su}, {Chen}, \& {Krucker}}]{2024SoPh..299..114R}
{Ryan}, D.~F., {Massa}, P., {Battaglia}, A.~F., {et~al.} 2024,
  \bibinfo{title}{{Triangulation of Hard X-Ray Sources in an X-Class Solar
  Flare with ASO-S/HXI and Solar Orbiter/STIX},} \solphys, 299, 114,
  \dodoi{10.1007/s11207-024-02341-8}

\bibitem[{D.~B. {Seaton} \& J.~M. {Darnel}(2018){Seaton} \&
  {Darnel}}]{2018ApJ...852L...9S}
{Seaton}, D.~B., \& {Darnel}, J.~M. 2018, \bibinfo{title}{{Observations of an
  Eruptive Solar Flare in the Extended EUV Solar Corona},} \apjl, 852, L9,
  \dodoi{10.3847/2041-8213/aaa28e}

\bibitem[{M.~Z. {Stiefel} {et~al.}(2025){Stiefel}, {Kuhar}, {Limousin},
  {Dickson}, {Volpara}, {Hurford}, \& {Krucker}}]{stiefel2025}
{Stiefel}, M.~Z., {Kuhar}, M., {Limousin}, O., {et~al.} 2025,
  \bibinfo{title}{{Using the STIX background detector as a proxy for GOES},}
  arXiv e-prints, arXiv:2501.03667, \dodoi{10.48550/arXiv.2501.03667}

\bibitem[{M.~Z. {Stiefel} {et~al.}(2023){Stiefel}, {Battaglia}, {Barczynski},
  {Collier}, {Volpara}, {Massa}, {Schwanitz}, {Tynelius}, {Harra}, \&
  {Krucker}}]{2023A&A...670A..89S}
{Stiefel}, M.~Z., {Battaglia}, A.~F., {Barczynski}, K., {et~al.} 2023,
  \bibinfo{title}{{Solar flare hard X-rays from the anchor points of an
  eruptive filament},} \aap, 670, A89, \dodoi{10.1051/0004-6361/202245044}

\bibitem[{A. {Thernisien}(2011){Thernisien}}]{2011ApJS..194...33T}
{Thernisien}, A. 2011, \bibinfo{title}{{Implementation of the Graduated
  Cylindrical Shell Model for the Three-dimensional Reconstruction of Coronal
  Mass Ejections},} \apjs, 194, 33, \dodoi{10.1088/0067-0049/194/2/33}

\bibitem[{A.~F.~R. {Thernisien} {et~al.}(2006){Thernisien}, {Howard}, \&
  {Vourlidas}}]{2006ApJ...652..763T}
{Thernisien}, A.~F.~R., {Howard}, R.~A., \& {Vourlidas}, A. 2006,
  \bibinfo{title}{{Modeling of Flux Rope Coronal Mass Ejections},} \apj, 652,
  763, \dodoi{10.1086/508254}

\bibitem[{S.~D. {Tun} \& A. {Vourlidas}(2013){Tun} \&
  {Vourlidas}}]{2013ApJ...766..130T}
{Tun}, S.~D., \& {Vourlidas}, A. 2013, \bibinfo{title}{{Derivation of the
  Magnetic Field in a Coronal Mass Ejection Core via Multi-frequency Radio
  Imaging},} \apj, 766, 130, \dodoi{10.1088/0004-637X/766/2/130}

\bibitem[{B. {Vr{\v{s}}nak} {et~al.}(2019){Vr{\v{s}}nak}, {Amerstorfer},
  {Dumbovi{\'c}}, {Leitner}, {Veronig}, {Temmer}, {M{\"o}stl}, {Amerstorfer},
  {Farrugia}, \& {Galvin}}]{2019ApJ...877...77V}
{Vr{\v{s}}nak}, B., {Amerstorfer}, T., {Dumbovi{\'c}}, M., {et~al.} 2019,
  \bibinfo{title}{{Heliospheric Evolution of Magnetic Clouds},} \apj, 877, 77,
  \dodoi{10.3847/1538-4357/ab190a}

\bibitem[{H. {Wang} {et~al.}(2015){Wang}, {Cao}, {Liu}, {Xu}, {Liu}, {Zeng},
  {Chae}, \& {Ji}}]{2015NatCo...6.7008W}
{Wang}, H., {Cao}, W., {Liu}, C., {et~al.} 2015, \bibinfo{title}{{Witnessing
  magnetic twist with high-resolution observation from the 1.6-m New Solar
  Telescope},} Nature Communications, 6, 7008, \dodoi{10.1038/ncomms8008}

\bibitem[{M.~J. {West} {et~al.}(2023){West}, {Seaton}, {Wexler}, {Raymond},
  {Del Zanna}, {Rivera}, {Kobelski}, {Chen}, {DeForest}, {Golub}, {Caspi},
  {Gilly}, {Kooi}, {Meyer}, {Alterman}, {Alzate}, {Andretta}, {Auch{\`e}re},
  {Banerjee}, {Berghmans}, {Chamberlin}, {Chitta}, {Downs}, {Giordano},
  {Harra}, {Higginson}, {Howard}, {Kumar}, {Mason}, {Mason}, {Morton},
  {Nykyri}, {Patel}, {Rachmeler}, {Reardon}, {Reeves}, {Savage}, {Thompson},
  {Van Kooten}, {Viall}, {Vourlidas}, \& {Zhukov}}]{West2023}
{West}, M.~J., {Seaton}, D.~B., {Wexler}, D.~B., {et~al.} 2023,
  \bibinfo{title}{{Defining the Middle Corona},} \solphys, 298, 78,
  \dodoi{10.1007/s11207-023-02170-1}

\bibitem[{F. {Yu} {et~al.}(2023){Yu}, {Zhao}, {Su}, {Zhu}, {Guo}, {Shen}, \&
  {Li}}]{2023ApJ...951...54Y}
{Yu}, F., {Zhao}, J., {Su}, Y., {et~al.} 2023, \bibinfo{title}{{Magnetic Field
  Extrapolation in Active Region Well Comparable to Observations in Multiple
  Layers},} \apj, 951, 54, \dodoi{10.3847/1538-4357/acd112}

\bibitem[{J. {Zhang} {et~al.}(2001){Zhang}, {Dere}, {Howard}, {Kundu}, \&
  {White}}]{2001ApJ...559..452Z}
{Zhang}, J., {Dere}, K.~P., {Howard}, R.~A., {Kundu}, M.~R., \& {White}, S.~M.
  2001, \bibinfo{title}{{On the Temporal Relationship between Coronal Mass
  Ejections and Flares},} \apj, 559, 452, \dodoi{10.1086/322405}

\bibitem[{Z. {Zhong} {et~al.}(2021){Zhong}, {Guo}, \&
  {Ding}}]{2021NatCo..12.2734Z}
{Zhong}, Z., {Guo}, Y., \& {Ding}, M.~D. 2021, \bibinfo{title}{{The role of
  non-axisymmetry of magnetic flux rope in constraining solar eruptions},}
  Nature Communications, 12, 2734, \dodoi{10.1038/s41467-021-23037-8}

\end{thebibliography}
\bibliographystyle{aasjournal}
\end{document}